\documentclass[aps,prl,twocolumn,titlepage,nofootinbib]{revtex4}
\usepackage{graphicx,physics}
\usepackage{color}
\usepackage{dcolumn}
\usepackage{latexsym}

\usepackage[normalem]{ulem}
\usepackage{hyperref,amssymb}
\usepackage{url}
\usepackage{color,soul}
\usepackage{graphicx}
\usepackage{verbatim}
\usepackage{multirow}
\usepackage{amsmath}
\newcommand{\beq}{\begin{eqnarray}}
\newcommand{\eeq}{\end{eqnarray}}
\usepackage{mathrsfs}
\usepackage{float,soul}
\usepackage[dvipsnames]{xcolor}
\usepackage{mathtools}
\usepackage{slashed}
\usepackage{physics}	
\usepackage{graphicx}   
\usepackage{epstopdf}
\usepackage{subfigure}  
\usepackage{hyperref}   
\usepackage{bbold}
\usepackage{wasysym}
\usepackage{feynmp}
\usepackage{hyperref}
\hypersetup{colorlinks,
}

\usepackage[latin1]{inputenc}

\begin{document}

\title{\Large Dispersionless Flat Mode and Vibrational Anomaly in Active Brownian Vibrators Induced by String-like Dynamical Defects}
\author{Cunyuan Jiang$^{1,2,3}$}
\author{Zihan Zheng$^{1,4}$}
\author{Yangrui Chen$^{1,4}$}
\author{Matteo Baggioli$^{1,2,3}$}
\email{b.matteo@sjtu.edu.cn}
\author{Jie Zhang$^{1,4}$}
\email{jiezhang2012@sjtu.edu.cn }
\address{$^1$ School of Physics and Astronomy, Shanghai Jiao Tong University, Shanghai 200240, China}
\address{$^2$ Wilczek Quantum Center, School of Physics and Astronomy, Shanghai Jiao Tong University, Shanghai 200240, China}
\address{$^3$ Shanghai Research Center for Quantum Sciences, Shanghai 201315,China}
\address{$^4$ Institute of Natural Sciences, Shanghai Jiao Tong University, Shanghai 200240, China}

\begin{abstract}
In recent years, active Brownian particles have emerged as a prominent model system for comprehending the behaviors of active matter, wherein particles demonstrate self-propelled motion by harnessing energy from the surrounding environment. A fundamental objective of studying active matter is to elucidate the physical mechanisms underlying its collective behaviors. Drawing inspiration from advancements in molecular glasses, our study unveils a low-energy ``flat mode" within the transverse spectrum of active Brownian vibrators -- a nearly two-dimensional, bi-disperse granular assembly. We demonstrate that this collective excitation induces an anomalous excess in the vibrational density of states (VDOS) beyond the phononic Debye contribution. We characterize the properties of this flat mode by exploring the parameter space of our experimental system and tuning the packing fraction, the vibrational frequency, the particle size ratio and the mixture ratio. Additionally, we establish through empirical evidence that string-like dynamical defects, discerned via the spatial distribution of each particle's contribution to the reduced transverse VDOS, serve as the microscopic origin of the flat mode and its associated anomalies. 
\end{abstract}
\maketitle
\color{blue}\textit{Introduction} 
\color{black} -- Active Brownian particles have recently gained popularity as a modeling system for investigating microorganisms such as bacteria, cells, and microswimmers \cite{Marchetti-RevModPhys.85.1143, Cates-annurev:/content/journals/10.1146/annurev-conmatphys-031214-014710, xu-PhysRevE.88.032304, kummel2013circular,reichhardt2014absorbing,henkes2020dense, sprenger2020active, caprini2023flocking, broker2023orientation}. Unlike passive Brownian particles, active Brownian particles do not exist in thermal equilibrium; they exhibit self-propelled motion by harnessing energy from their surrounding environment \cite{Marchetti-RevModPhys.85.1143, Cates-annurev:/content/journals/10.1146/annurev-conmatphys-031214-014710}. Understanding the behavior of these active particles is crucial for deciphering processes like cell migration in intricate tissue and organ formation \cite{Marchetti-RevModPhys.85.1143}, as well as wound healing \cite{safferling2013wound}. Furthermore, a deeper understanding of active Brownian particles is imperative for the development of artificial micro- and nanorobots intended for future applications in drug delivery and environmental protection.

Chen et al. recently conducted a quasi two-dimensional (2D) experiment of uniformly driven active Brownian vibrators \cite{PhysRevE.106.L052903, chen2023anomalous}, where single-particle velocity and rotation follow Gaussian distributions with zero means\cite{PhysRevE.106.L052903}. The system consists of hard-disk-like particles with no inherent orientation or preferred directions, unlike previous studies \cite{kudrolli-PhysRevLett.100.058001, narayan2007long, dauchot-PhysRevLett.105.098001, scholz-PhysRevLett.118.198003}, resembling active Brownian particles from theory \cite{caprini2023flocking}. Unlike microswimmers in viscous media, the system's underdamped dynamics provide a platform for directly observing collective shear excitations in bi-disperse systems \cite{2403.08285} while avoiding crystallization issues. 

Gels, structural glasses, colloids, and granular assemblies, are commonplace. However, their properties remain poorly understood in contrast to ideal crystals \cite{chaikin2000principles}. These amorphous systems exhibit non-phononic degrees of freedom, collectively termed as `defects,' which contribute to the observed anomalies alongside crystal-like acoustic waves.
Within amorphous materials, the concept of `defects' manifests in various forms. For instance, they are recognized as two-level systems \cite{doi:10.1080/14786437208229210,Phillips1972} elucidating the low-temperature heat capacity \cite{PhysRevB.4.2029}, quasilocalized non-phononic modes \cite{10.1063/5.0069477} accounting for the boson peak (BP) excess in the vibrational density of states \cite{ramos2022low}, and as entities such as shear transformation zones \cite{ARGON197947}, Eshelby quadrupoles \cite{doi:10.1098/rspa.1959.0173}, or topological defects \cite{Baggioli2023}, providing insight into mechanical and plastic dynamics during deformation \cite{annurev:/content/journals/10.1146/annurev-conmatphys-062910-140452}.

Recent studies have highlighted the importance of string-like dynamical defects in materials such as supercooled liquids, heated crystals, quasi-crystals, and glasses. These defects, similar to vortex lines in superfluids \cite{kleinert1989gauge}, influence relaxational and dynamical properties \cite{PhysRevLett.80.2338,10.1063/1.4878502,10.1063/5.0039162,10.1063/1.4918807,10.1063/1.4769267,zhao2024quasicrystals,doi:10.1021/acs.jpcb.9b09468}. Simulations reveal these quasi-1D transverse entities as likely contributors to nonphononic modes and the microscopic origin of the boson peak (BP) in glassy materials \cite{PhysRevB.101.174311,Hu2022,PhysRevResearch.5.023055,d1,jiang2023stringlet,jiang2024phonons,10.1063/5.0197386}. Hu and Tanaka's simulations \cite{Hu2022,PhysRevResearch.5.023055} show a strong link between these excitations and a ``flat boson peak mode," observed experimentally in 2D amorphous materials \cite{Tomterud2023}. Despite experimental evidence in amorphous systems \cite{PhysRevE.81.041301,doi:10.1073/pnas.0900227106,doi:10.1073/pnas.1101858108}, direct observation of these defects in amorphous active Brownian particles remains absent. Enhanced understanding of vibrational and collective dynamics in these particles could shed light on active matter mechanics and hydrodynamics \cite{Marchetti-RevModPhys.85.1143, ramaswamy2010mechanics}.

\color{blue}Experimental setup and data analysis \color{black} -- Our experimental system consists in a horizontal layer of granular particles driven vertically by a sinusoidal oscillation with a fixed frequency $f$ and amplitude of $0.062$ mm induced by an electromagnetic shaker. We utilize bi-disperse particles with a size ratio of disk diameters of $d_s/d_l$ and a number ratio of $N_s/N_l$ for small and large particle respectively. Different packing fractions $\phi$ can be achieved. The motion of the particles, labeled ``granular Brownian vibrators", is recorded with a Basler CCD camera (acA2040-180kc) at $40$ frames/s for at least an hour. We refer to Refs. \cite{PhysRevE.106.L052903,2403.08285} for more details about the experimental setup.

The analysis of the experimental data is based on the dynamical matrix $D_{ij}$ defined from the displacement correlation matrix. By diagonalizing the dynamical matrix, one obtains the eigenfrequencies $\omega_i$ and the corresponding eigenvector fields $\boldsymbol{u}_i$. Using standard methods \cite{binder2011glassy}, longitudinal ($L$) and transverse ($T$) dynamical structure factors $S_{L,T} (\boldsymbol{k},\omega)$ are extracted. The vibrational density of states (VDOS) can be obtained by $g(\omega)=\sum_i \delta_{\omega,\omega_i}$, or equivalently $g(\omega)_{L,T}=\int S_{L,T} (\boldsymbol{k},\omega) d\boldsymbol{k}$, showing perfect agreement. Finally, the particle level VDOS is given by $g_{L,T}(\boldsymbol{r}_j,\omega)=\sum_{i} |\boldsymbol{u}_{i,L,T} (\boldsymbol{r}_{j})|^2 \delta_{\omega,\omega_i}$ where $\boldsymbol{u}_{i,L,T}(\boldsymbol{r}_j)$ is the longitudinal or transverse eigenvector field of $j$th particle under eigenfrequency $\omega_i$ \cite{Hu2022,PhysRevResearch.5.023055}. Further details about the computation of all these quantities can be found in the Supplementary Information (SI).

\begin{figure}[ht]
    \centering
    \includegraphics[width=0.6\linewidth]{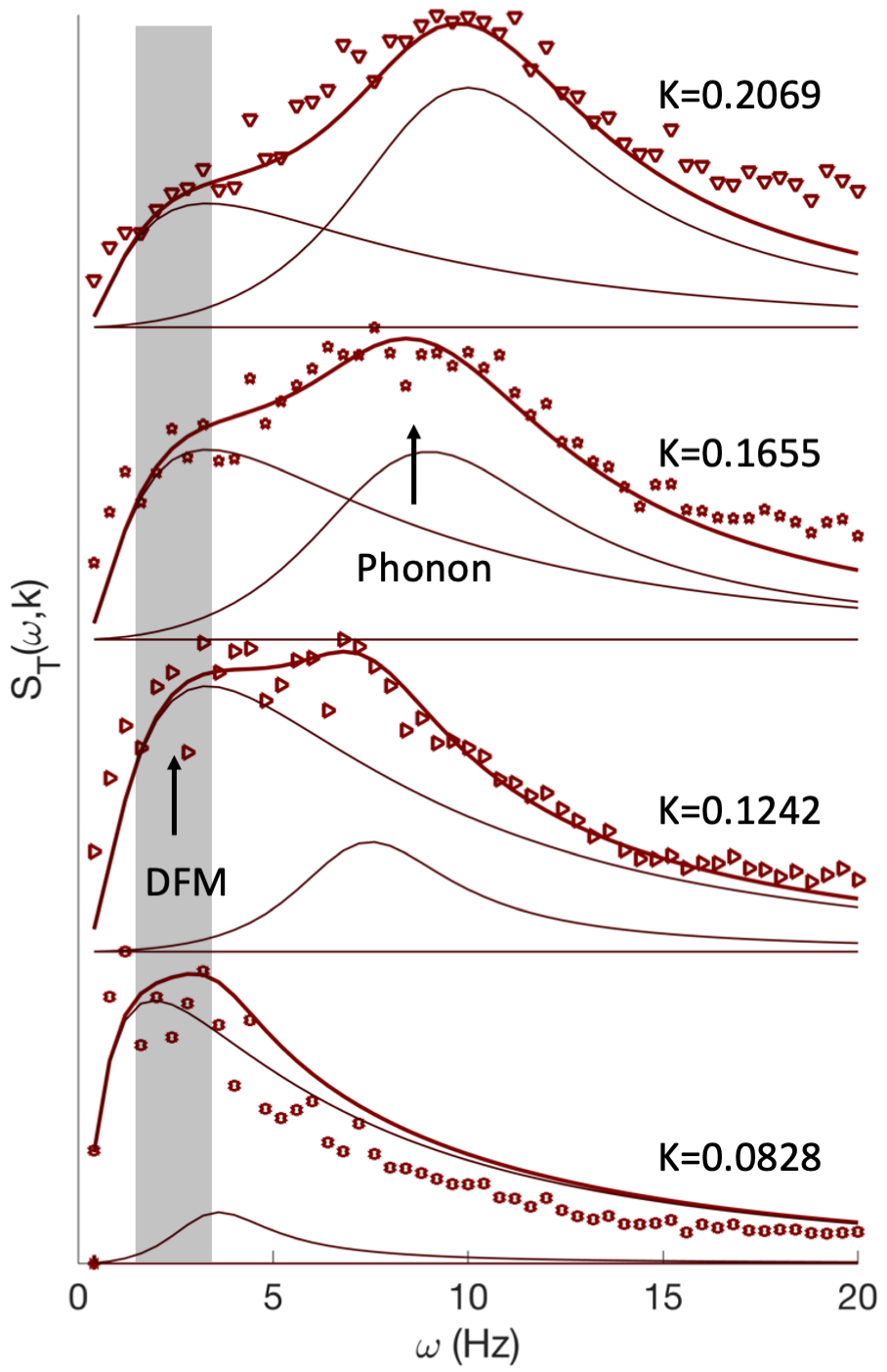}%
    \caption{Transverse dynamical structure factor $S_T(\omega,k)$. The red symbols are the experimental data. The solid lines represent the fit using Eq.\eqref{susu}. The gray vertical region indicates the energy of the dispersionless flat mode (DFM) plus/minus its uncertainty.}
    \label{figdsflines}
\end{figure}

\color{blue}Dynamical response and vibrational anomalies \color{black} -- As benchmark example, we consider a system with $\phi=0.822$, $f=100$ Hz, $d_s/d_l=1/1.4$ and $N_s/N_l=2/1$. The transverse dynamical structure factor is shown as a function of frequency and for different wave-vectors in Fig.\ref{figdsflines}. The experimental data are visualized with red symbols. The thick red lines are the results of a fit with the following function,
\begin{equation}\label{susu}
    S_{T}(\omega,k) \propto \dfrac{\omega^2}{(\omega^2-\Omega_{T}^2)^2+ \omega^2 \Gamma_T^2} +S_{\text{extra}}(\omega).
\end{equation}
with
    $S_{\text{extra}}(\omega)=\dfrac{1}{\omega \sigma \sqrt{2 \pi}} \mathrm{exp}\left( -\dfrac{(\mathrm{ln} \omega-\mu)^2}{2 \sigma^2} \right)$,
while the thin lines are the two contributions in Eq. \eqref{susu}.
The first term corresponds to a damped harmonic oscillator with linewidth $\Gamma_T(k)$ and characteristic frequency $\Omega_T(k)$. The second term represents a log-normal function \cite{MALINOVSKY199163} that describes additional low-energy modes as commonly utilized in amorphous systems \cite{Hu2022,PhysRevResearch.5.023055}. 

The results of Fig.\ref{figdsflines} outline the presence of two distinct excitations, captured by the two terms in Eq.\eqref{susu}. The first mode corresponds to propagating shear waves (transverse phonons) with a speed of $v_T=71.7$ d$_{s}$/s (in unit of the small particle diameter $d_s$), that is extracted from the $k$ dependence of $\Omega_T$ (see SI for details). At this specific value of the packing fraction, the system is in a dense liquid state and the collective shear waves present a small gap in wave-vector \cite{2403.08285}, known as $k$-gap \cite{BAGGIOLI20201}, that emerges nevertheless at smaller wave-vector and is not shown in Fig.\ref{figdsflines}. See SI for other packing fractions.

The second excitation appears at frequencies below those of the transverse phonon and corresponds to the peak whose position is highlighted by the vertical gray area. Interestingly, we find that the energy of this collective mode does not depend on the wave-vector $k$. Hence, we label this excitation the ``\textit{dispersionless flat mode}'' (DFM). As the wave-vector becomes larger, the intensity of the peak associated to the DFM becomes weaker with respect to that of the concomitant transverse phonon.

Importantly, a DFM can only be found in the transverse dynamical structure factor, indicating its transverse nature. As demonstrated in the SI, the longitudinal dynamical structure factor $S_L(\omega,k)$ displays a single peak, corresponding to a longitudinal phonon mode with linear dispersion relation, $\Omega_L(k)=v_L k$, and speed $v_L=116.9$ d$_s$/s \cite{2403.08285}. 

\begin{figure}[htb]
    \centering
    \includegraphics[width=0.8\linewidth]{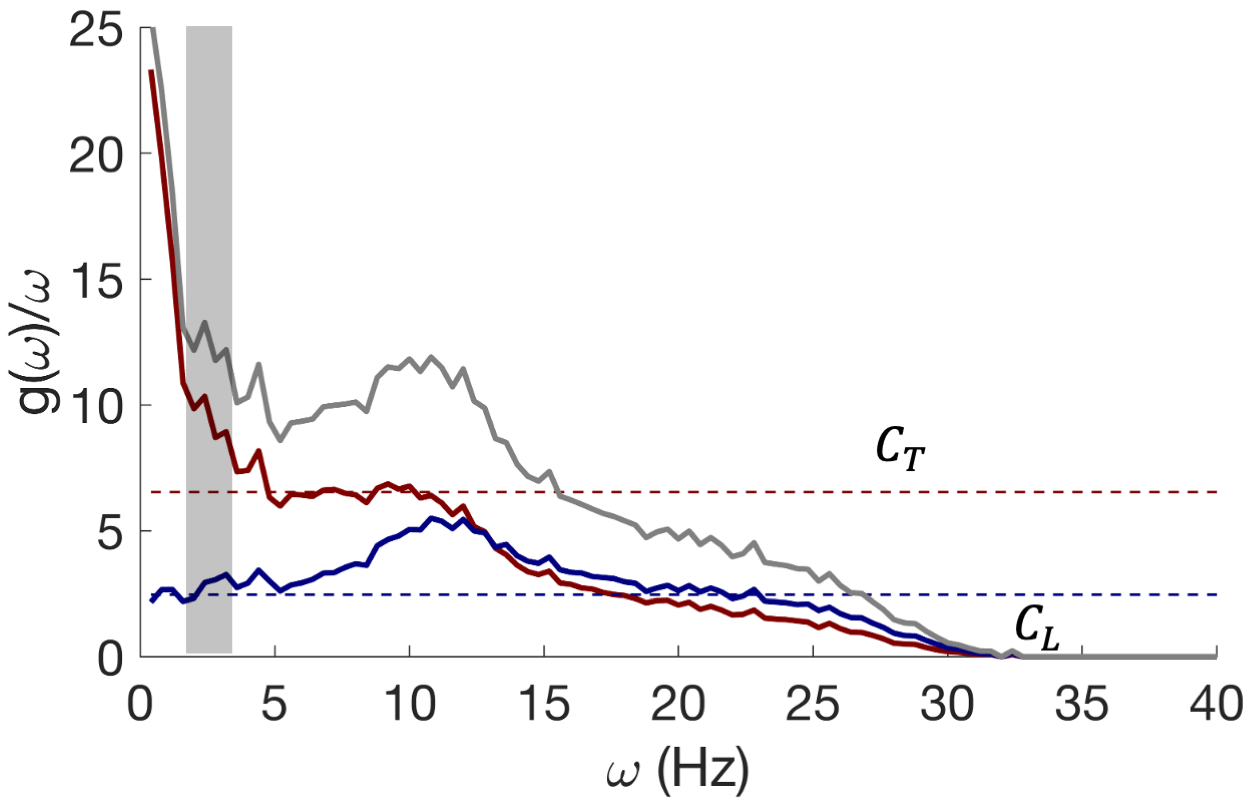}%
    \caption{Reduced VDOS for $\phi=0.822$. The blue, red, and gray colors correspond to the longitudinal, transverse, and total VDOS. The vertical gray area indicates the energy of the dispersionless flat mode observed in Fig.\ref{figdsflines}. The dashed lines denote the position of the longitudinal and transverse Debye levels, $C_{L,T}$, determined from the corresponding sound speeds.}
    \label{figvdos}
\end{figure}

To confirm the existence of the DFM and study its impact, we move to the analysis of the VDOS, $g(\omega)$. For better visualization, we normalize the VDOS by the 2D Debye scaling $g_{\text{Debye}}\propto \omega$. Our experimental results are shown in Fig.\ref{figvdos}, where the blue, red and gray lines correspond respectively to the longitudinal, transverse and total density of states.

At low frequency, the longitudinal part of $g(\omega)/\omega$ is approximately constant, indicating the validity of the Debye approximation, and consistent with a linearly dispersing longitudinal phonon. It then shows a pseudo-Van Hove peak at approximately $12$ Hz, that agrees with the energy at which $\Omega_L(k)$ flattens (see SI), and then decays at higher frequencies, vanishing above $\approx 30$ Hz.

The transverse component of $g(\omega)/\omega$ (red curve in Fig.\ref{figvdos}), presents a more complex behavior that can be divided in three different regions. Below $1$ Hz, the transverse VDOS decays sharply indicating the presence of a strong diffusive component that can be also seen from a large nonzero value of $g_T(\omega)$ at zero frequency (see inset of Fig.\ref{figvdos}). Within $5 \sim 12$ Hz, the transverse $g(\omega)/\omega$ is almost flat with a much weaker pseudo-Van Hove peak than the longitudinal one. This indicates that the transverse dynamics are more overdamped than the longitudinal ones. In order to test the validity of Debye theory, we focus on the low frequency range in which the transverse and longitudinal reduced VDOS display an approximate plateau. We mark the corresponding plateaus with horizontal dashed lines in Fig.\ref{figvdos}. We verify that the ratio of the values of the two plateaus, $C_L /C_T$, agrees well with the prediction of the Debye model, $v_T^2/v_L^2$ where the speeds are extracted from the dispersion relations obtained from the dynamical structure factor.

Interestingly, within $1\sim5$ Hz, we observe an excess anomaly in the transverse $g(\omega)/\omega$, aligning with the DFM observed in Fig.\ref{figdsflines}, whose structure is similar to the BP observed in several amorphous materials \cite{ramos2022low}.

\begin{figure}[ht!]
    \centering
    \includegraphics[width=\linewidth]{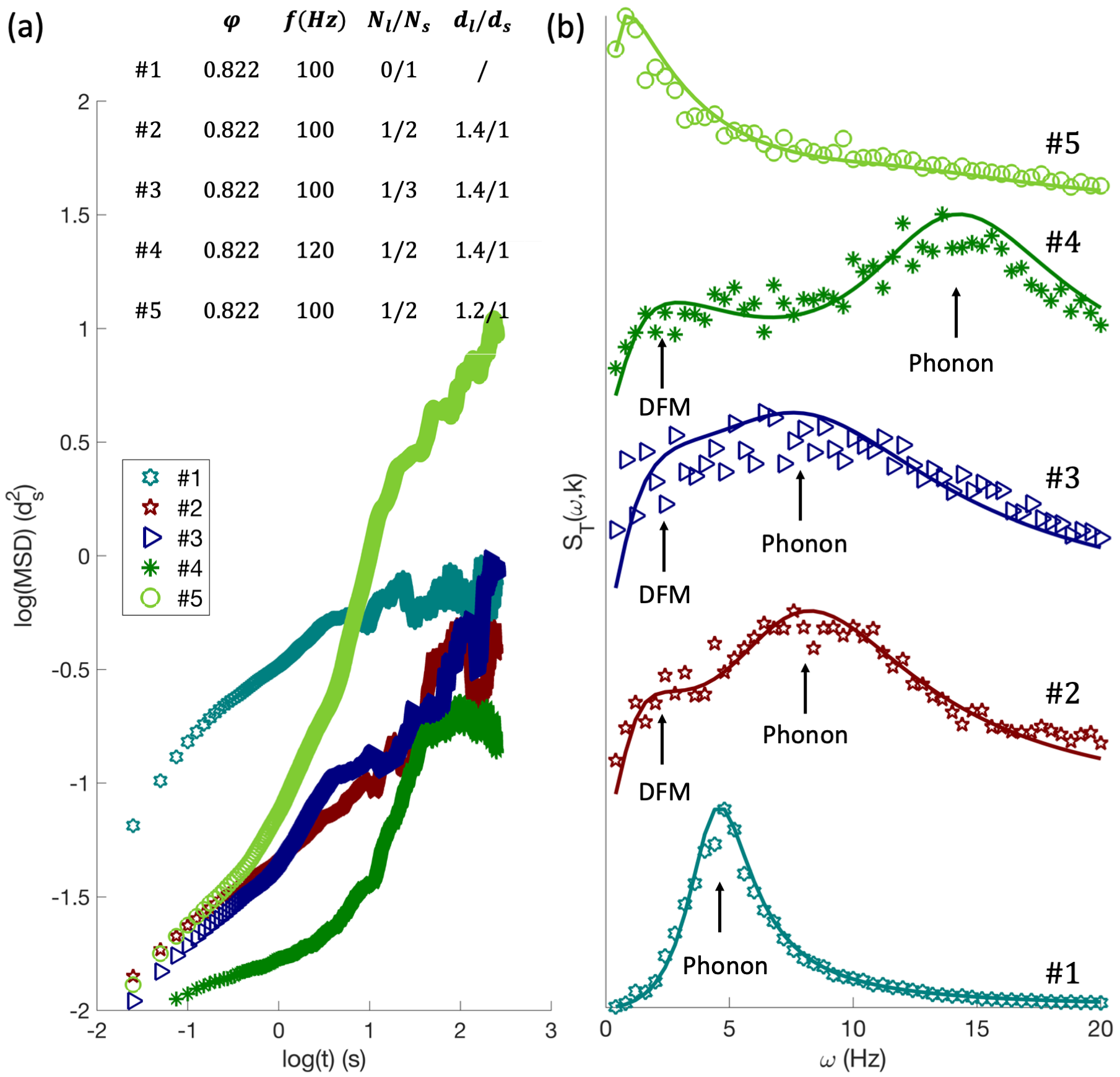}
    \caption{\textbf{(a)} Mean square displacement for different experimental setups. The inset lists the parameters corresponding to each dataset, where $\phi$ is packing fraction, $f$ driving frequency, $N_l / N_s$ large and small particles mixture ratio and $d_l / d_s$ their size ratio. \textbf{(b)} Transverse dynamical structure factor at $k=0.1655 d_s^{-1}$ for each dataset.}
    \label{fig:4}
\end{figure}

\color{blue}Dynamics of the dispersionless flat mode \color{black} -- In order to test under which experimental conditions the DFM appears, we performed four additional experiments by keeping the packing fraction constant and moving independently the other three parameters $f$, $N_s/N_l$, $d_s/d_l$. The values of these parameters for the different experimental datasets can be found in the inset of Fig.~\ref{fig:4}. Dataset $\#2$ corresponds to the benchmark case presented in Figs.~\ref{figdsflines}-\ref{figvdos}.

First, in dataset $\#1$ we show the results for a mono-disperse crystalline packing (see image in SI). As one might expect, at late times ($t>10$ s), the particle mean square displacement (MSD) reaches a constant plateau. On the other hand, the transverse structure factor is well approximated by a single damped harmonic oscillator contribution, corresponding to a dispersing transverse phonon. The DOS (shown in SI) is also perfectly compatible with the Debye model. 
\\

In dataset $\#3$, we have increased the relative number of small particles. The MSD shows a faster late time increase and the features in $ S_{T}(\omega,k)$ become more incoherent. In particular, despite two peaks are still evident (and their frequencies unchanged), their width becomes very large at the point that they almost merge. This corresponds to a situation where damping and friction are very large and the excitations, both the transverse phonon and the DFM, become overdamped.

In dataset $\#4$, we increased the vibration frequency. Since the acceleration of the EM shaker is kept constant, this corresponds to decrease the amplitude of the vibrations. As a result, dataset $\#4$ displays a more solid-like behavior compared to our original sample $\#2$. In particular, the MSD is much smaller and the late-time plateau more evident. Moreover, the frequency of the second phononic peak in $ S_{T}(\omega,k)$ is larger, indicating that the material is stiffer. Also, as shown by the green dark line in Fig.~\ref{fig:4}(b), the low energy DFM peak is sharper but its frequency remains approximately unchanged.

Finally, dataset $\#5$ corresponds to a smaller size ratio between large and small particles. The late-time MSD grows quickly with time following a linear scaling (diffusion), indicating that under these conditions the experimental system is more akin a dilute liquid. The results for $ S_{T}(\omega,k)$ (and the DOS, see SI) confirm this picture. $ S_{T}(\omega,k)$ does not present any sharp transverse phonon peak but rather only a very weak low-frequency excess corresponding to a softened DFM that is barely visible. 

The varying behaviors of different samples shown in Fig.\ref{fig:4} can be observed directly in the videos available in the SI, and it is also confirmed by the value of $S_T(k,\omega=0)$ that increases with the degree of ``fluidity''. This analysis provides a consistent framework for the nature of the DFM and its characteristics. In summary, we can conclude that the DFM results from the disordered structure of our granular system since it is absent in the crystalline packing. Moreover, we observe a natural tendency of this flat mode to become more overdamped and soft by moving towards configurations in the parameter space where the system is ``\textit{less solid}'' and more akin to a fluid. We notice that this behavior is similar to what happens with the BP in structural glasses upon increasing the temperature and moving well above the glass transition temperature (see \textit{e.g.} \cite{PhysRevLett.92.245508,10.1063/1.2360275}), a behavior that has recently been rationalized using stringlet excitations \cite{jiang2023stringlet}.

Finally, in the SI, we also show the dynamical structure factor for lower packing fraction values ($\phi=0.811,0.695$) to verify that decreasing $\phi$, the characteristics of our experimental system become more liquid-like. At lower $\phi$, the DFM peak softens and becomes more damped, eventually disappearing, hidden by the large low-frequency signal related to the liquid-like relaxational mechanisms. This confirms that $\phi$ plays an analogous role as the inverse of an effective temperature \cite{2403.08285}, melting the solid and the related flat mode.

\begin{figure}[htb]
    \centering
    \includegraphics[width=0.8\linewidth]{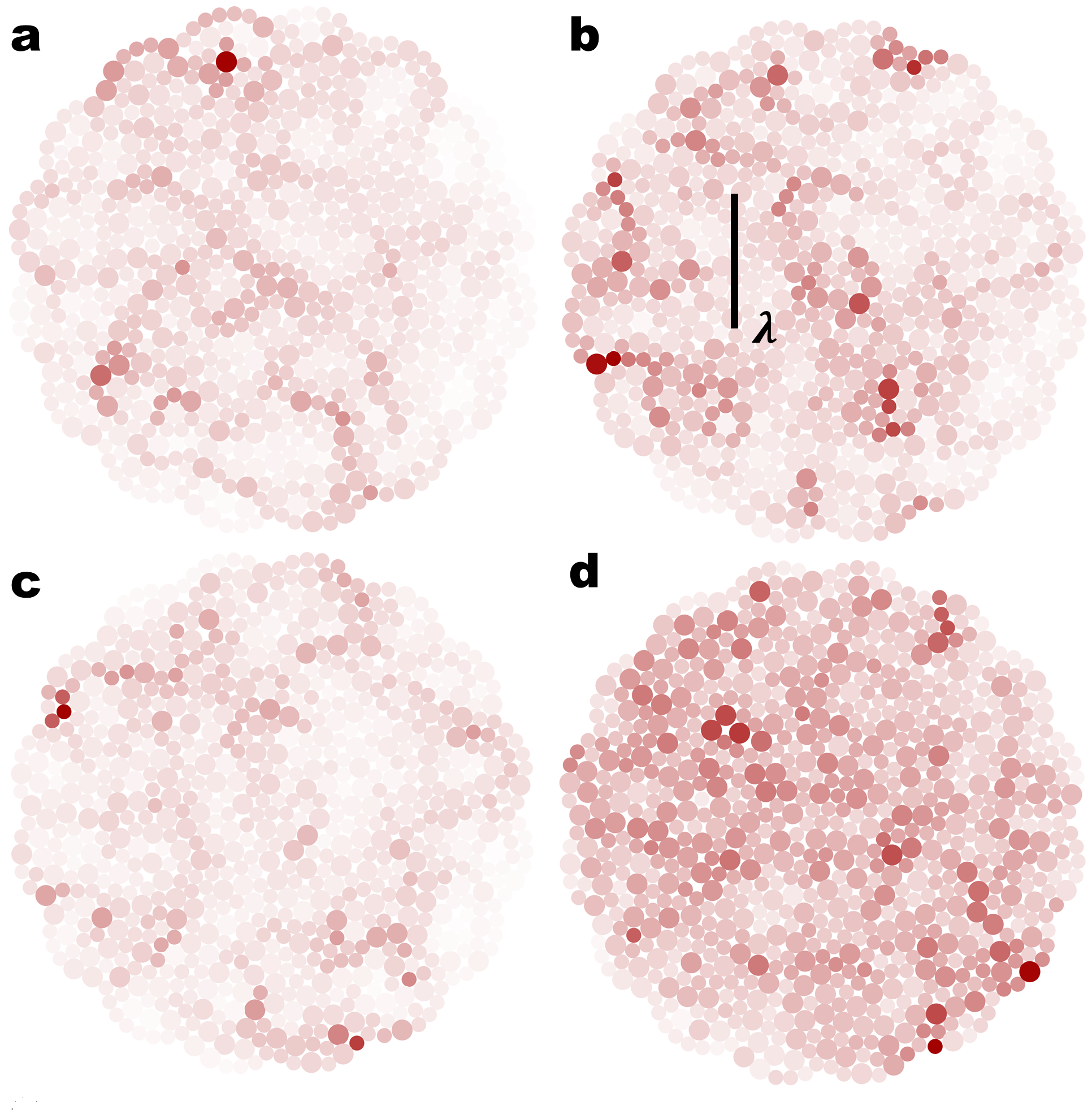}%
    \caption{Spatial distributions of transverse particle-level reduced VDOS at frequencies $0.4,1.2,2,8,$ Hz from panel \textbf{(a)} to \textbf{(d)}. A darker color indicates a more significant contribution. The thick black line in panel (b) shows the extent of the average length of string-like dynamical defects, approximately $12 d_s$.}
    \label{figplvdos}
\end{figure}

\color{blue}String-like dynamical defects \color{black} -- To reveal the microscopic origin of the DFM, we consider the particle-level reduced transverse VDOS, characterizing individual particle contribution at a given frequency.
The results are shown for the benchmark dataset $\#2$ in Fig.\ref{figplvdos} for six different frequency values of $\omega$, from $0.4$ to $12$ Hz. Darker red color indicates a more significant contribution to the reduced transverse VDOS at that frequency.

Panels (a-c) are for frequencies around the characteristic energy of the DFM, as indicated with the vertical gray region in Figs.\ref{figdsflines}-\ref{figvdos}. Panel (b) lies precisely at the frequency of the flat mode corresponding to the BP energy. In this range of frequencies, 1D string-like structures emerge, indicating particles' heterogeneous dynamics and tendency to form collective excitations of approximately 1D nature. String-like dynamical defects are most substantial at the flat mode frequency in panel (b).

However, for a higher frequency, as shown in panel (d), the spatial distribution of the particle-level VDOS becomes more uniform, and no evident string-like structure is observed anymore.
On the other hand, as shown explicitly in SI, the longitudinal particle-level VDOS does not display any string-like structure. This confirms the DFM's transverse nature and its relation to string-like excitations.

\color{black}

To provide a stronger connection between the string-like defect structures, the BP anomaly in the VDOS and the properties of the DFM, we follow the idea of associating a length scale with the BP introduced by Hong et al. \cite{PhysRevE.83.061508,HONG2011351} and Kalampounias et al. \cite{10.1063/1.2360275} on a heuristic basis (see also \cite{MALINOVSKY1988111,MALINOVSKY1986757,https://doi.org/10.1002/pssb.2220640120}). We then define a cooperative length scale, $l= \alpha v_T/ \omega_{BP}= \alpha v_T /\omega_{\text{DFM}}$, where $\alpha$ is an unknown $\mathcal{O}(1)$ coefficient. Note that previous works connected the scale of dynamic heterogeneities $\xi$ to the BP frequency, $\xi \approx c v_T/\omega_{BP}$, where $c$ is a constant found to be between $1/2$ and $1$ \cite{https://doi.org/10.1002/pssc.200777584,Duval_1990,PhysRevLett.69.1540,Elliott_1992,Elliott_1992,PhysRevLett.100.137402,PhysRevB.66.174205}. Hence, we also empirically assume $1/2<\alpha<1$. Using $\omega_{\text{DFM}}=\omega_{BP}\approx 2 $ Hz, we find $17 d_s<l< 35 \,d_s$. This value is compatible (in order of magnitude) with the average length of the string-like structures $\lambda \approx 12 \,d_s$ observed in Fig.\ref{figplvdos}, despite being slightly larger, confirming the microscopic origin of the flat mode might be the revealed string-like dynamical defects. 

This result agrees qualitatively with recent theoretical studies \cite{jiang2023stringlet,jiang2024phonons}. At zero temperature, or equivalently in the absence of damping, the length-scale $l$ extracted from the BP and the average string-like defects length $\lambda$ were found to match to a good approximation. In overdamped dynamics, the BP-like anomaly is softened by damping mechanisms, resulting in a longer length scale, $l>\lambda$. As observed in glass-forming liquids above the glass transition temperature \cite{jiang2023stringlet}, in that case $l$ is slightly larger than $\lambda$, consistent with our findings. We then ascribe this small deviation of $\l$ with respect $\lambda$ to anharmonic and frictional effects.

\color{blue}\textit{Conclusions}\color{black} -- We conducted an experiment utilizing a quasi-2D disordered active Brownian particle system. Our investigation unveiled a robust correlation among a collective dispersionless flat mode, the boson peak anomaly in the vibrational density of states, and the emergence of string-like dynamical defects. Our findings affirm that string-like structures, predominantly of a transverse nature, may underlie the low-energy vibrational anomalies observed in active Brownian particles. This hypothesis was initially proposed in molecular glasses through simulations \cite{Hu2022}, and the existence of the boson peak aligns with a dispersionless localized mode, as evidenced experimentally in 2D silica \cite{Tomterud2023}. Our discoveries indicate striking parallels in the low-energy collective excitations of disordered active Brownian particles and molecular glasses, despite the myriad of distinctions between these systems.

\color{blue}{\it Acknowledgments} \color{black} --  We would like to thank Jack Douglas, Yunjiang Wang, Wensheng Xu, Haibin Yu, Peng Tan and Massimo Pica Ciamarra for numerous discussions about string-like objects and their role in amorphous systems. CJ and MB acknowledge the support of the Shanghai Municipal Science and Technology Major Project (Grant No.2019SHZDZX01). MB acknowledges the support of the sponsorship from the Yangyang Development Fund.
ZZ, YC, and JZ acknowledge the support of the NSFC (No. 11974238 and No. 12274291) and the Shanghai Municipal Education Commission Innovation Program under No. 2021-01-07-00-02-E00138. ZZ, YC, and JZ also acknowledge the support from the Shanghai Jiao Tong University Student Innovation Center.

\newpage
\onecolumngrid
\appendix 
\clearpage
\renewcommand\thefigure{S\arabic{figure}}    
\setcounter{figure}{0} 
\renewcommand{\theequation}{S\arabic{equation}}
\setcounter{equation}{0}
\renewcommand{\thesubsection}{SM\arabic{subsection}}
\section*{\Large Supplementary Information}
\textbf{In this Supplementary Information (SI), we provide details on the methods used and further analysis on our experimental system.}

\subsection*{Methods}\label{meth}
We define the displacement correlation matrix $\boldsymbol{C}$ \cite{PhysRevLett.104.248305,PhysRevLett.105.025501},
\begin{equation}
    C_{i j}=\langle n(t)_i n(t)_j \rangle_T,
\end{equation}
where $n_i (t)$ is the displacement of $i$th degree of freedom at time $t$, and $T$ the time window on which the average $\langle \cdot \rangle$ is performed. The dynamical matrix can then be calculated as,
\begin{equation}
    D_{i j}=\dfrac{\beta\, C^{-1}_{i j}}{\sqrt{m_i m_j}},
\end{equation}
where $m_i$ is the mass of the $i$th degree of freedom, and $\beta$ is a dimensionful parameter which is fixed using the same procedure outlined in \cite{2403.08285}. By diagonalizing the matrix $\boldsymbol{D}$, one obtains the eigenvalues $\kappa_i$ and the eigenfrequencies $\omega_i = \sqrt{\kappa_i}$. The eigenvector fields $\boldsymbol{u}$ are then defined by solving the eigenvalue problem, $\boldsymbol{D}\boldsymbol{u}= \omega^2 \boldsymbol{u}$. From the eigenvector field $\boldsymbol{u}_i$, longitudinal ($L$) and transverse ($T$) dynamical structure factor are extracted,
\begin{equation}
    S_{L,T} (\boldsymbol{k},\omega) \propto \dfrac{1}{k^2} \left| \sum_i F_{(L,T),i} (\boldsymbol{k}) \delta_{\omega,\omega_i} \right|^2,
\end{equation}
with
\begin{align}
    &F_{L,i} (\boldsymbol{k})=\boldsymbol{k} \cdot \sum_j \boldsymbol{u}_{i} (\boldsymbol{r}_j) e^{-i \boldsymbol{k} \cdot \boldsymbol{r}_j},\\
   & F_{T,i} (\boldsymbol{k})=\boldsymbol{z} \cdot \boldsymbol{k} \times \sum_j \boldsymbol{u}_{i} (\boldsymbol{r}_j) e^{-i \boldsymbol{k} \cdot \boldsymbol{r}_j}.
\end{align}
Here, $\boldsymbol{z}$ is the vector perpendicular to the two-dimensional experiment platform, while the index $j$ runs over all particles and $\boldsymbol{r}_j$ is the position of $j$th particle. 

We clarify that $\boldsymbol{k}$ is a vector with length $k=|\boldsymbol{k}|$ and direction from $0$ to $2 \pi$. Also, in our notation, the Fourier transform is given by
\begin{equation}
    \mathcal{F}[\boldsymbol{k}]=\int f(\boldsymbol{r}) e^{- i 2 \pi \boldsymbol{k} \cdot \boldsymbol{r}} d\boldsymbol{r}.
\end{equation}
\subsection*{Extended analysis}

\subsubsection*{Longitudinal dynamics}
\begin{figure}[htb]
    \centering
    \includegraphics[width=0.3\linewidth]{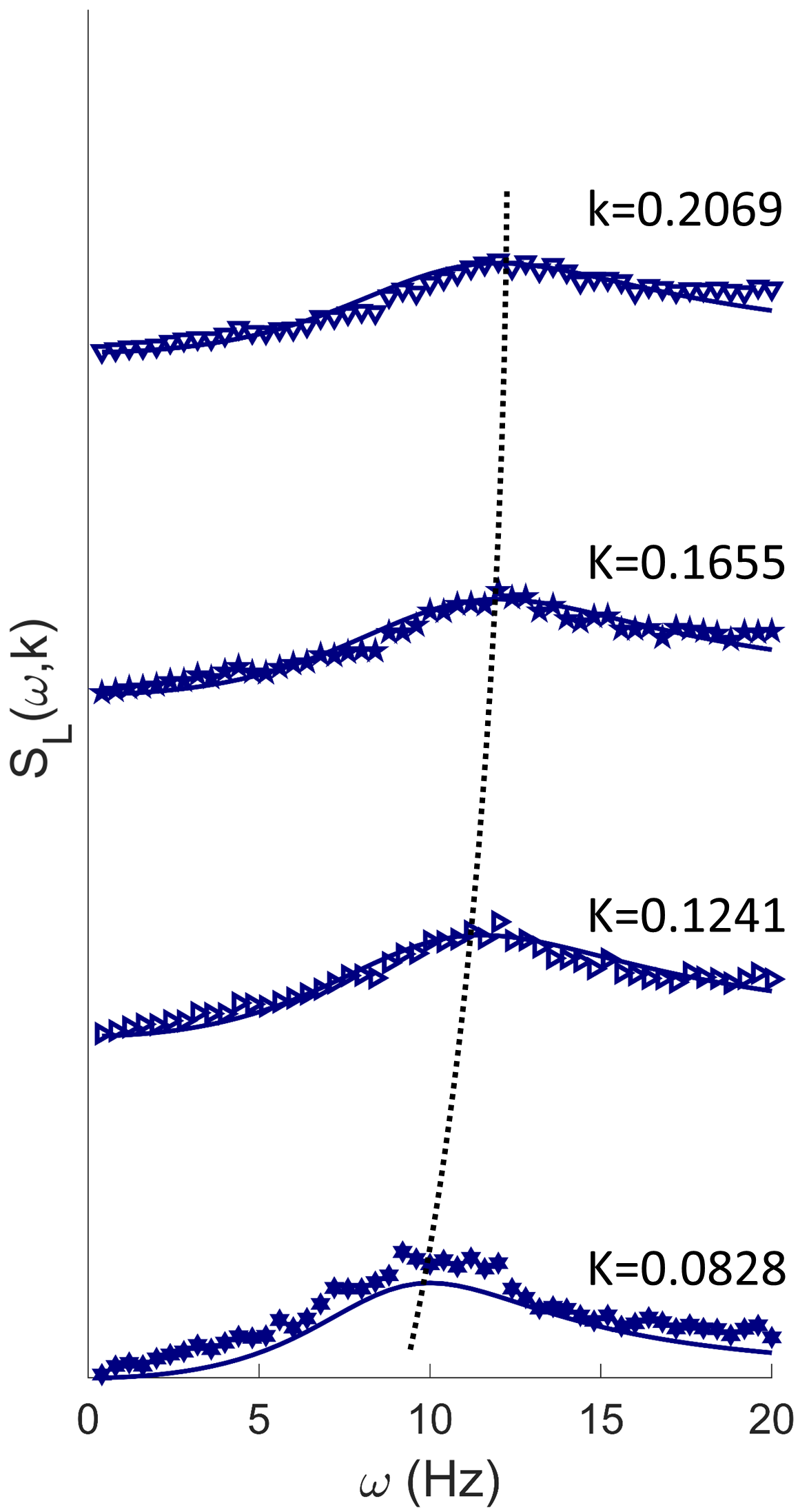}%
    \caption{The longitudinal dynamical structure factor $S_{L}(\omega,k)$ as a function of the frequency for different wave-vectors (labeled in figure). The solid lines indicate the fit to the damped harmonic oscillator function, Eq. \eqref{susi}. The black dashed line indicates the dispersion of the longitudinal mode $\Omega_L(k)$.}
    \label{figdsflinesl}
\end{figure}

The longitudinal dynamical structure factor for dataset $\#2$ in the main text ($\phi=0.822$) as a function of frequency and for different wave-vectors is shown in Fig.\ref{figdsflinesl}. It can be fitted by a single damped harmonic oscillator function,
\begin{equation}\label{susi}
    S_{L}(\omega,k) \propto \dfrac{\omega^2}{(\omega^2-\Omega_{L}^2)^2+ \omega^2 \Gamma_L^2}
\end{equation}
that corresponds to an attenuated longitudinal phonon. The dispersion relation of this mode, indicated with a black dashed line in Fig.\ref{figdsflinesl}, is well approximated by a linear function $\Omega_L=v_L k$, with $v_L=116.9$ d$_s$/s.

\begin{figure}[htb]
    \centering
    \includegraphics[width=0.5\linewidth]{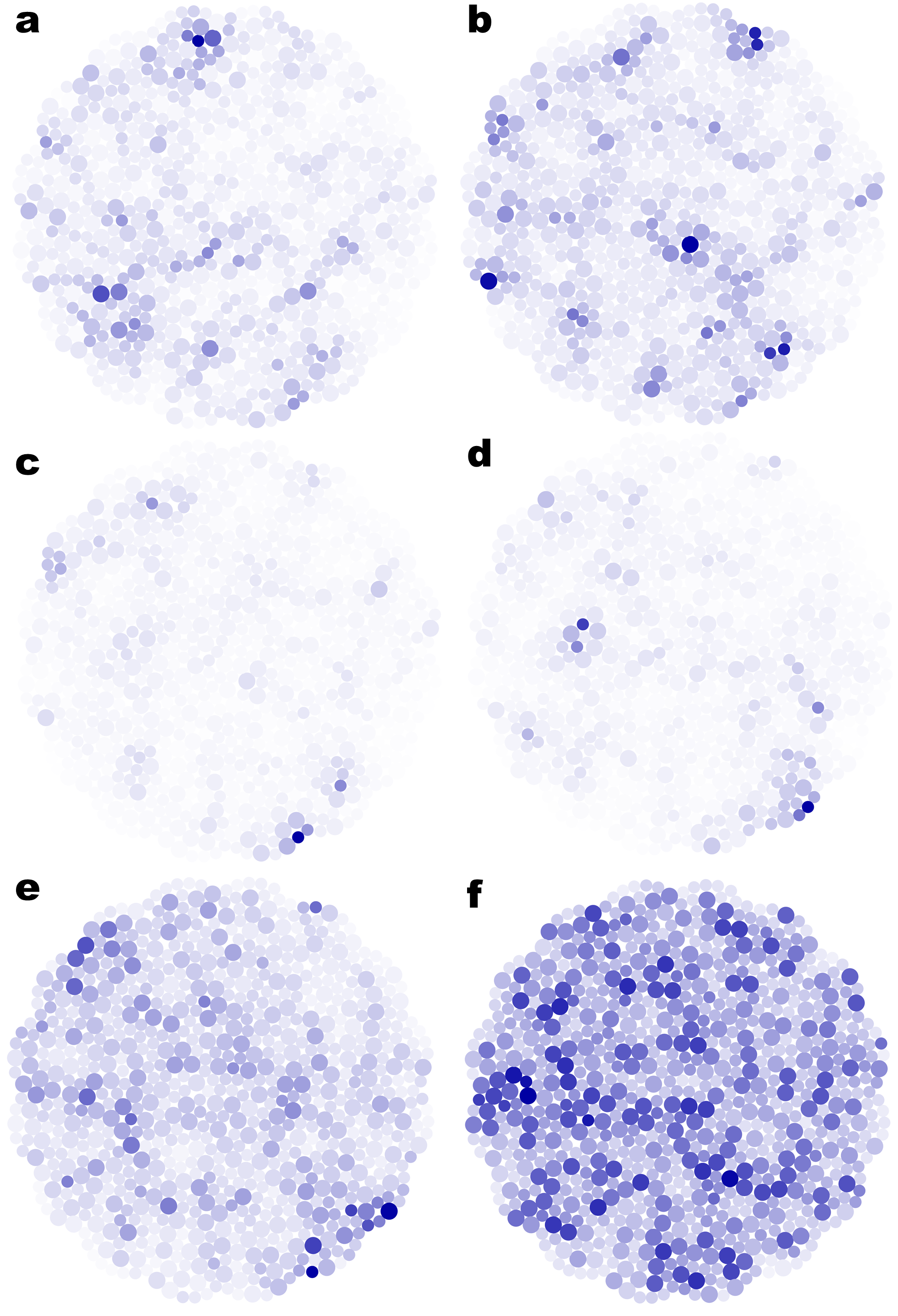}%
    \caption{The distribution of longitudinal particle level VDOS with frequency $0.4,1.2,2,4,8,12$ Hz from panel \textbf{(a)} to \textbf{(f)}. The darker the color, the more contribution the particle makes to VDOS at the selected frequency.}
    \label{figplvdosl}
\end{figure}

In Fig.\ref{figplvdosl}, we present the particle level longitudinal vibrational density of states (VDOS). Contrary to the transverse component discussed in the main text, we do no observe any clear evidence of string-like dynamical defects.

\subsubsection*{Transverse phonon dispersion}
\begin{figure}
    \centering
    \includegraphics[width=0.4\linewidth]{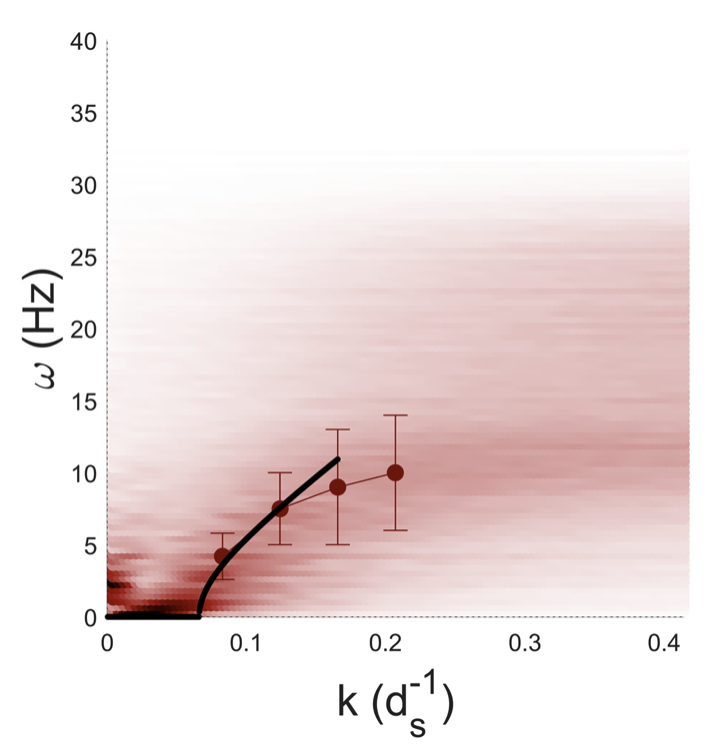}
    \caption{Dispersion relation of the transverse phonon for dataset $\#2$ with $\phi=0.822$. The thick line is the fit to equation \eqref{gigi}. The symbols correspond to the location of the peak in the dynamical structure factor, and the error bars show their width.  The background color is the value of the transverse dynamical structure factor $S_T(\omega,k)$.}
    \label{ohoh}
\end{figure}

Even at very large values of the packing fraction $\phi$, the dispersion relation of the transverse phonon displays a small wave-vector gap that grows by reducing $\phi$ \cite{2403.08285}. Therefore, in order to extract the speed of transverse sound we use the following formula,
\begin{equation}\label{gigi}
    \Omega_T(k)=v_T \sqrt{k^2-k_g^2},
\end{equation}
where $\Omega_T$ is the phonon energy extracted from the dynamical structure factor. In Fig.\ref{ohoh}, we show an example of this procedure for dataset $\#2$ corresponding to the highest packing fraction $\phi=0.822$. The value of $v_T$ reported in the main text is obtained in this way.
\subsubsection*{Low packing fraction data}
\begin{figure}[ht!]
    \centering
    \includegraphics[height=0.6\linewidth]{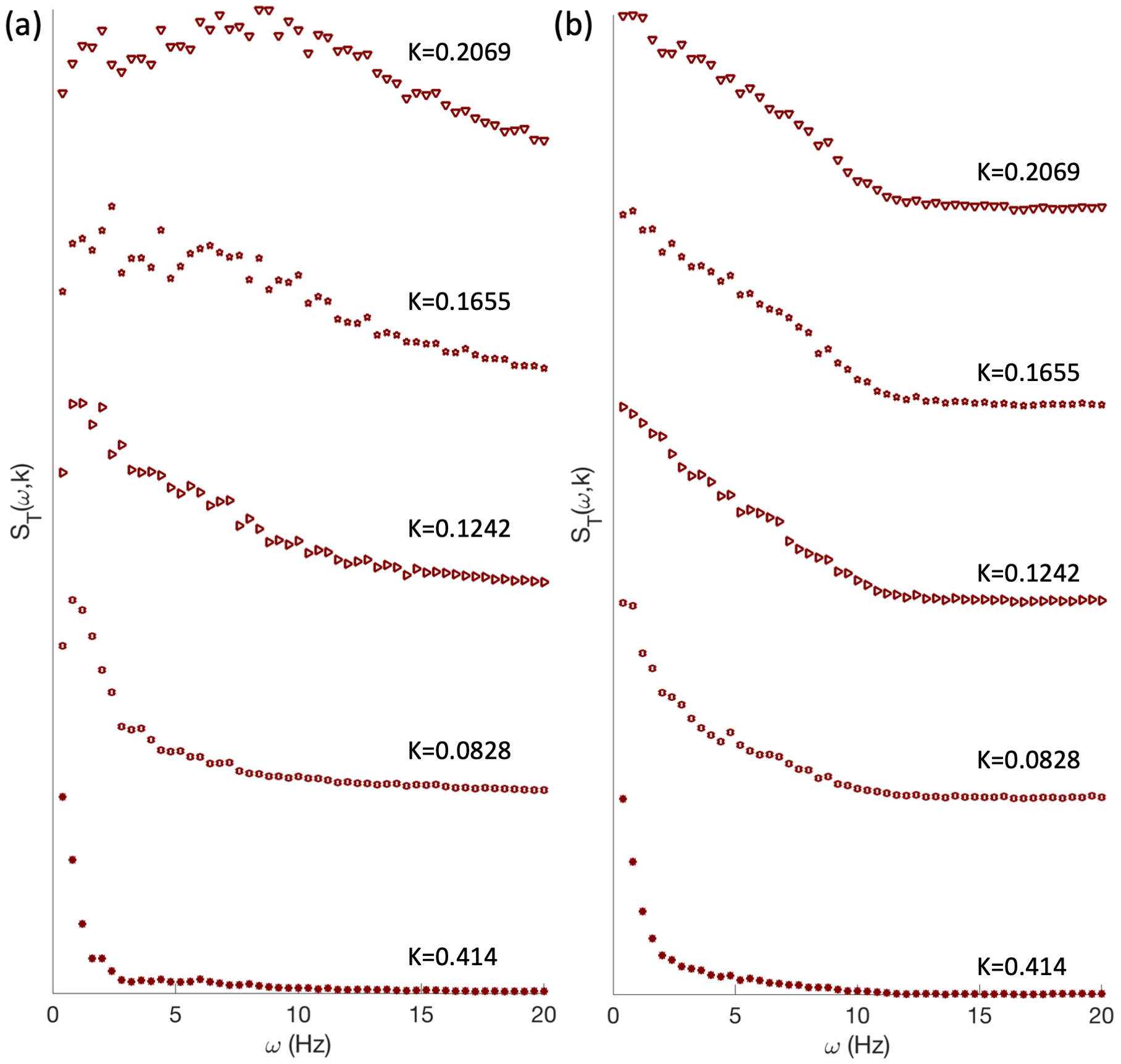}
    \caption{Transverse dynamical structure factor as a function of the frequency $\omega$ for different wave-vectors $k$ for $\phi=0.811$ \textbf{(a)} and $\phi=0.695$ \textbf{(b)}.}
    \label{f2}
\end{figure}

In Fig.~\ref{f2} we show the dynamical structure factor for systems with lower packing fraction. First, in panel (a) we show the results for $\phi=0.811$ which is slightly lower than the packing fraction value used in the main text. From there, it is evident that both the transverse phonon and the flat mode are overdamped at low wave-vector. By increasing the wave-vector, the two modes re-emerge but the width of corresponding peaks remains still quite large. In panel (b) of Fig.\ref{f2}, we show the same quantity for a much lower value of the packing fraction, $\phi=0.695$. At such small packing fraction, no clear peak is observed in $S_T(\omega,k)$ that, on the contrary, decays monotonically. These findings are consistent with the idea \cite{2403.08285} that decreasing the packing fraction is equivalent to melting our amorphous solid system and moving towards a more fluid-like (or even gas-like at very small $\phi$) phase.

\subsubsection*{VDOS for different experimental datasets}
\begin{figure}[htb]
    \centering
    \includegraphics[width=0.7\linewidth]{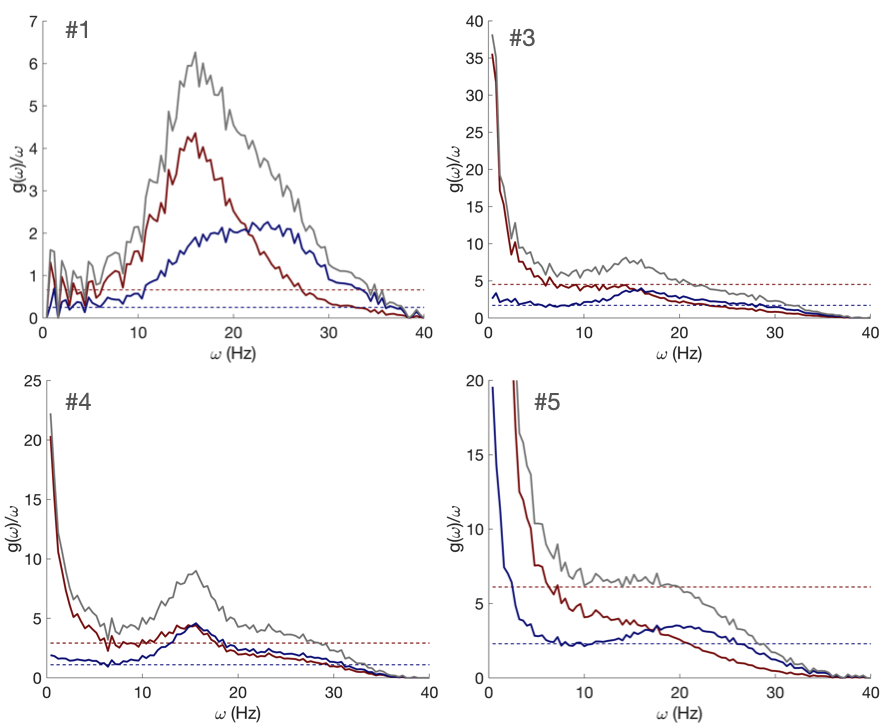}%
    \caption{The reduced VDOS for the datasets whose dynamics are presented in Fig.~\ref{fig:4}. The VDOS for the benchmark case $\#2$ is shown in Fig.~\ref{figvdos} in the main text. The blue, red, and gray colors correspond to the longitudinal, transverse, and total VDOS. The dashed lines denote the position of the longitudinal and transverse Debye levels, $C_{L,T}$, determined from the corresponding sound speeds.}
    \label{figalldos}
\end{figure}

In Fig.\ref{figalldos}, we present the experimental data for the VDOS for different datasets discussed in the main text. Dataset $\#1$ corresponds to a crystalline packing of mono-disperse particles. Both longitudinal and transverse VDOS exhibit consistency with the Debye law at small frequencies within the experimental noise and display two sharp Van-Hove singularities at approximately $18$ Hz and $25$ Hz for transverse and longitudinal phonons, respectively. 

Dataset $\#4$ represents the most ``solid" of the amorphous packings. The longitudinal VDOS follows the Debye law at low frequency, while the transverse VDOS displays a large $\omega \approx 0$ relaxational signal and a small bump at intermediate frequencies corresponding to the flat mode. Both components show quite pronounced pseudo-Van-Hove singularity at around $17$ Hz. 

Moving towards a more liquid-like regime, dataset $\#3$ exhibits a stronger diffusive $\omega \approx 0$ signal (at least twice larger) and the pseudo-Van-Hove singularities become overdamped and almost disappear. 

Finally, dataset $\#5$ is the most liquid-like, where the zero frequency signal has become very large, and the longitudinal part of the VDOS no longer follows the Debye law at low frequency. Also, the pseudo-Van-Hove singularities are entirely washed out. The characteristics in the VDOS reported in Fig.\ref{figalldos} confirm the results presented in the main text.

\subsubsection*{Images of the different packings}
\begin{figure}[htb]
    \centering
    \includegraphics[width=0.7\linewidth]{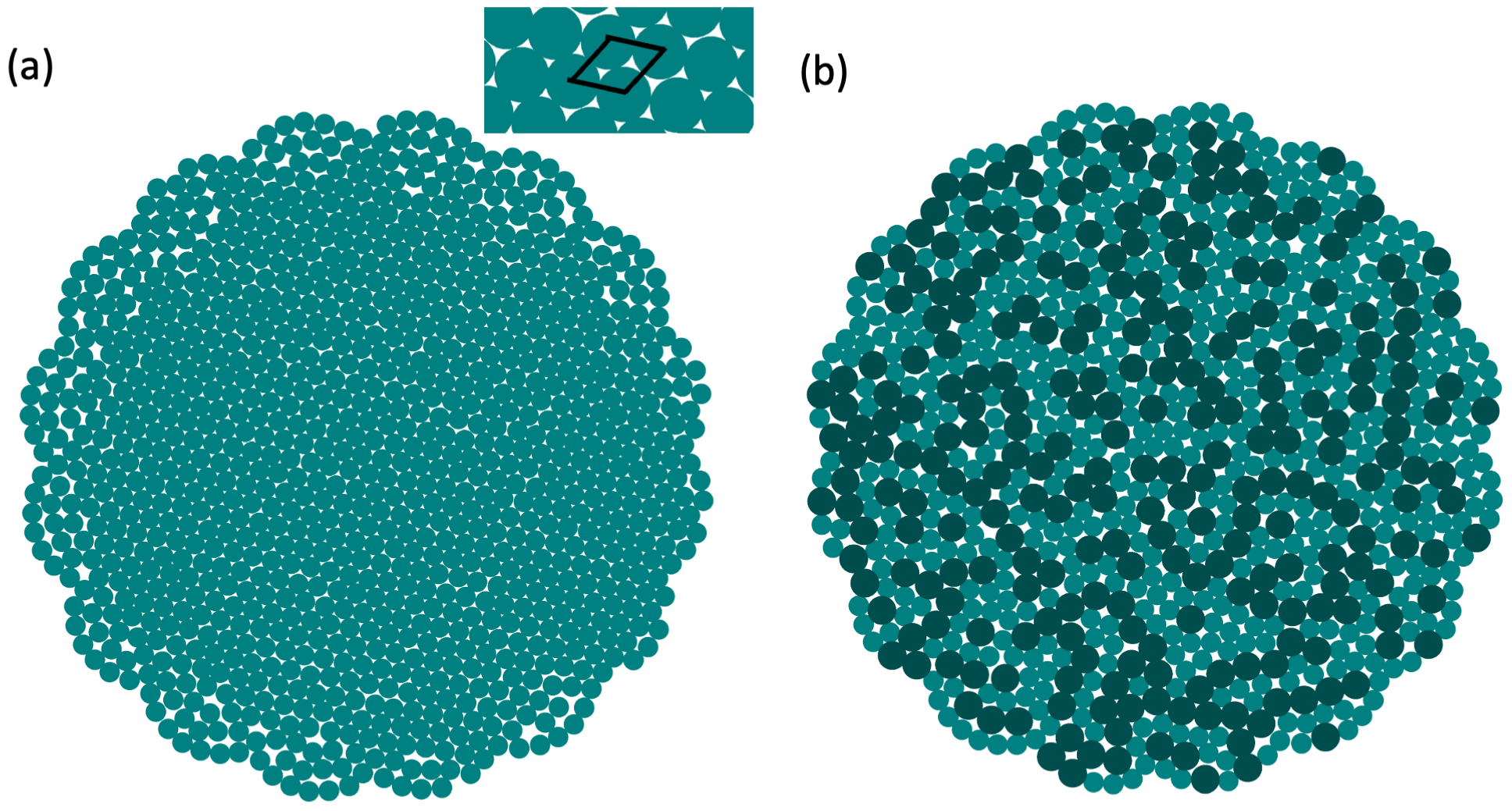}%
    \caption{Images of the crystalline mono-disperse packing \textbf{(a)} and the amorphous bi-disperse packing \textbf{(b)} for $\phi=0.822$.}
    \label{figstructure}
\end{figure}

In Fig.\ref{figstructure}, we show images of the two experimental setups corresponding to the crystalline mono-disperse packing and the amorphous bi-disperse packing with $\phi=0.822$.

\subsubsection*{Real-time videos of microscopic particle motion}
Attached to this manuscript, the Reader can find real-time videos of the particle motion corresponding to the different datasets discussed in the main text. In particular, the video files \color{blue}\textit{video $\#n$} \color{black} correspond to the $n$-th dataset.
Direct visualization of the microscopic particle motion confirms the more solid-like or liquid-like nature of the different datasets. Particle re-arrangements are more evident and frequent in the more liquid-like samples. All videos record $500s$ of experimental runs.

\begin{thebibliography}{65}%
\makeatletter
\providecommand \@ifxundefined [1]{%
 \@ifx{#1\undefined}
}%
\providecommand \@ifnum [1]{%
 \ifnum #1\expandafter \@firstoftwo
 \else \expandafter \@secondoftwo
 \fi
}%
\providecommand \@ifx [1]{%
 \ifx #1\expandafter \@firstoftwo
 \else \expandafter \@secondoftwo
 \fi
}%
\providecommand \natexlab [1]{#1}%
\providecommand \enquote  [1]{``#1''}%
\providecommand \bibnamefont  [1]{#1}%
\providecommand \bibfnamefont [1]{#1}%
\providecommand \citenamefont [1]{#1}%
\providecommand \href@noop [0]{\@secondoftwo}%
\providecommand \href [0]{\begingroup \@sanitize@url \@href}%
\providecommand \@href[1]{\@@startlink{#1}\@@href}%
\providecommand \@@href[1]{\endgroup#1\@@endlink}%
\providecommand \@sanitize@url [0]{\catcode `\\12\catcode `\$12\catcode `\&12\catcode `\#12\catcode `\^12\catcode `\_12\catcode `\%12\relax}%
\providecommand \@@startlink[1]{}%
\providecommand \@@endlink[0]{}%
\providecommand \url  [0]{\begingroup\@sanitize@url \@url }%
\providecommand \@url [1]{\endgroup\@href {#1}{\urlprefix }}%
\providecommand \urlprefix  [0]{URL }%
\providecommand \Eprint [0]{\href }%
\providecommand \doibase [0]{http://dx.doi.org/}%
\providecommand \selectlanguage [0]{\@gobble}%
\providecommand \bibinfo  [0]{\@secondoftwo}%
\providecommand \bibfield  [0]{\@secondoftwo}%
\providecommand \translation [1]{[#1]}%
\providecommand \BibitemOpen [0]{}%
\providecommand \bibitemStop [0]{}%
\providecommand \bibitemNoStop [0]{.\EOS\space}%
\providecommand \EOS [0]{\spacefactor3000\relax}%
\providecommand \BibitemShut  [1]{\csname bibitem#1\endcsname}%
\let\auto@bib@innerbib\@empty
\bibitem [{\citenamefont {Marchetti}\ \emph {et~al.}(2013)\citenamefont {Marchetti}, \citenamefont {Joanny}, \citenamefont {Ramaswamy}, \citenamefont {Liverpool}, \citenamefont {Prost}, \citenamefont {Rao},\ and\ \citenamefont {Simha}}]{Marchetti-RevModPhys.85.1143}%
  \BibitemOpen
  \bibfield  {author} {\bibinfo {author} {\bibfnamefont {M.~C.}\ \bibnamefont {Marchetti}}, \bibinfo {author} {\bibfnamefont {J.~F.}\ \bibnamefont {Joanny}}, \bibinfo {author} {\bibfnamefont {S.}~\bibnamefont {Ramaswamy}}, \bibinfo {author} {\bibfnamefont {T.~B.}\ \bibnamefont {Liverpool}}, \bibinfo {author} {\bibfnamefont {J.}~\bibnamefont {Prost}}, \bibinfo {author} {\bibfnamefont {M.}~\bibnamefont {Rao}}, \ and\ \bibinfo {author} {\bibfnamefont {R.~A.}\ \bibnamefont {Simha}},\ }\href {\doibase 10.1103/RevModPhys.85.1143} {\bibfield  {journal} {\bibinfo  {journal} {Rev. Mod. Phys.}\ }\textbf {\bibinfo {volume} {85}},\ \bibinfo {pages} {1143} (\bibinfo {year} {2013})}\BibitemShut {NoStop}%
\bibitem [{\citenamefont {Cates}\ and\ \citenamefont {Tailleur}(2015)}]{Cates-annurev:/content/journals/10.1146/annurev-conmatphys-031214-014710}%
  \BibitemOpen
  \bibfield  {author} {\bibinfo {author} {\bibfnamefont {M.~E.}\ \bibnamefont {Cates}}\ and\ \bibinfo {author} {\bibfnamefont {J.}~\bibnamefont {Tailleur}},\ }\href {\doibase https://doi.org/10.1146/annurev-conmatphys-031214-014710} {\bibfield  {journal} {\bibinfo  {journal} {Annual Review of Condensed Matter Physics}\ }\textbf {\bibinfo {volume} {6}},\ \bibinfo {pages} {219} (\bibinfo {year} {2015})}\BibitemShut {NoStop}%
\bibitem [{\citenamefont {Zheng}\ \emph {et~al.}(2013)\citenamefont {Zheng}, \citenamefont {ten Hagen}, \citenamefont {Kaiser}, \citenamefont {Wu}, \citenamefont {Cui}, \citenamefont {Silber-Li},\ and\ \citenamefont {L\"owen}}]{xu-PhysRevE.88.032304}%
  \BibitemOpen
  \bibfield  {author} {\bibinfo {author} {\bibfnamefont {X.}~\bibnamefont {Zheng}}, \bibinfo {author} {\bibfnamefont {B.}~\bibnamefont {ten Hagen}}, \bibinfo {author} {\bibfnamefont {A.}~\bibnamefont {Kaiser}}, \bibinfo {author} {\bibfnamefont {M.}~\bibnamefont {Wu}}, \bibinfo {author} {\bibfnamefont {H.}~\bibnamefont {Cui}}, \bibinfo {author} {\bibfnamefont {Z.}~\bibnamefont {Silber-Li}}, \ and\ \bibinfo {author} {\bibfnamefont {H.}~\bibnamefont {L\"owen}},\ }\href {\doibase 10.1103/PhysRevE.88.032304} {\bibfield  {journal} {\bibinfo  {journal} {Phys. Rev. E}\ }\textbf {\bibinfo {volume} {88}},\ \bibinfo {pages} {032304} (\bibinfo {year} {2013})}\BibitemShut {NoStop}%
\bibitem [{\citenamefont {K{\"u}mmel}\ \emph {et~al.}(2013)\citenamefont {K{\"u}mmel}, \citenamefont {Ten~Hagen}, \citenamefont {Wittkowski}, \citenamefont {Buttinoni}, \citenamefont {Eichhorn}, \citenamefont {Volpe}, \citenamefont {L{\"o}wen},\ and\ \citenamefont {Bechinger}}]{kummel2013circular}%
  \BibitemOpen
  \bibfield  {author} {\bibinfo {author} {\bibfnamefont {F.}~\bibnamefont {K{\"u}mmel}}, \bibinfo {author} {\bibfnamefont {B.}~\bibnamefont {Ten~Hagen}}, \bibinfo {author} {\bibfnamefont {R.}~\bibnamefont {Wittkowski}}, \bibinfo {author} {\bibfnamefont {I.}~\bibnamefont {Buttinoni}}, \bibinfo {author} {\bibfnamefont {R.}~\bibnamefont {Eichhorn}}, \bibinfo {author} {\bibfnamefont {G.}~\bibnamefont {Volpe}}, \bibinfo {author} {\bibfnamefont {H.}~\bibnamefont {L{\"o}wen}}, \ and\ \bibinfo {author} {\bibfnamefont {C.}~\bibnamefont {Bechinger}},\ }\href@noop {} {\bibfield  {journal} {\bibinfo  {journal} {Physical review letters}\ }\textbf {\bibinfo {volume} {110}},\ \bibinfo {pages} {198302} (\bibinfo {year} {2013})}\BibitemShut {NoStop}%
\bibitem [{\citenamefont {Reichhardt}\ and\ \citenamefont {Reichhardt}(2014)}]{reichhardt2014absorbing}%
  \BibitemOpen
  \bibfield  {author} {\bibinfo {author} {\bibfnamefont {C.}~\bibnamefont {Reichhardt}}\ and\ \bibinfo {author} {\bibfnamefont {C.~J.~O.}\ \bibnamefont {Reichhardt}},\ }\href@noop {} {\bibfield  {journal} {\bibinfo  {journal} {Soft Matter}\ }\textbf {\bibinfo {volume} {10}},\ \bibinfo {pages} {7502} (\bibinfo {year} {2014})}\BibitemShut {NoStop}%
\bibitem [{\citenamefont {Henkes}\ \emph {et~al.}(2020)\citenamefont {Henkes}, \citenamefont {Kostanjevec}, \citenamefont {Collinson}, \citenamefont {Sknepnek},\ and\ \citenamefont {Bertin}}]{henkes2020dense}%
  \BibitemOpen
  \bibfield  {author} {\bibinfo {author} {\bibfnamefont {S.}~\bibnamefont {Henkes}}, \bibinfo {author} {\bibfnamefont {K.}~\bibnamefont {Kostanjevec}}, \bibinfo {author} {\bibfnamefont {J.~M.}\ \bibnamefont {Collinson}}, \bibinfo {author} {\bibfnamefont {R.}~\bibnamefont {Sknepnek}}, \ and\ \bibinfo {author} {\bibfnamefont {E.}~\bibnamefont {Bertin}},\ }\href@noop {} {\bibfield  {journal} {\bibinfo  {journal} {Nature communications}\ }\textbf {\bibinfo {volume} {11}},\ \bibinfo {pages} {1405} (\bibinfo {year} {2020})}\BibitemShut {NoStop}%
\bibitem [{\citenamefont {Sprenger}\ \emph {et~al.}(2020)\citenamefont {Sprenger}, \citenamefont {Fernandez-Rodriguez}, \citenamefont {Alvarez}, \citenamefont {Isa}, \citenamefont {Wittkowski},\ and\ \citenamefont {Lowen}}]{sprenger2020active}%
  \BibitemOpen
  \bibfield  {author} {\bibinfo {author} {\bibfnamefont {A.~R.}\ \bibnamefont {Sprenger}}, \bibinfo {author} {\bibfnamefont {M.~A.}\ \bibnamefont {Fernandez-Rodriguez}}, \bibinfo {author} {\bibfnamefont {L.}~\bibnamefont {Alvarez}}, \bibinfo {author} {\bibfnamefont {L.}~\bibnamefont {Isa}}, \bibinfo {author} {\bibfnamefont {R.}~\bibnamefont {Wittkowski}}, \ and\ \bibinfo {author} {\bibfnamefont {H.}~\bibnamefont {Lowen}},\ }\href@noop {} {\bibfield  {journal} {\bibinfo  {journal} {Langmuir}\ }\textbf {\bibinfo {volume} {36}},\ \bibinfo {pages} {7066} (\bibinfo {year} {2020})}\BibitemShut {NoStop}%
\bibitem [{\citenamefont {Caprini}\ and\ \citenamefont {L{\"o}wen}(2023)}]{caprini2023flocking}%
  \BibitemOpen
  \bibfield  {author} {\bibinfo {author} {\bibfnamefont {L.}~\bibnamefont {Caprini}}\ and\ \bibinfo {author} {\bibfnamefont {H.}~\bibnamefont {L{\"o}wen}},\ }\href@noop {} {\bibfield  {journal} {\bibinfo  {journal} {Physical Review Letters}\ }\textbf {\bibinfo {volume} {130}},\ \bibinfo {pages} {148202} (\bibinfo {year} {2023})}\BibitemShut {NoStop}%
\bibitem [{\citenamefont {Br{\"o}ker}\ \emph {et~al.}(2023)\citenamefont {Br{\"o}ker}, \citenamefont {Bickmann}, \citenamefont {Te~Vrugt}, \citenamefont {Cates},\ and\ \citenamefont {Wittkowski}}]{broker2023orientation}%
  \BibitemOpen
  \bibfield  {author} {\bibinfo {author} {\bibfnamefont {S.}~\bibnamefont {Br{\"o}ker}}, \bibinfo {author} {\bibfnamefont {J.}~\bibnamefont {Bickmann}}, \bibinfo {author} {\bibfnamefont {M.}~\bibnamefont {Te~Vrugt}}, \bibinfo {author} {\bibfnamefont {M.~E.}\ \bibnamefont {Cates}}, \ and\ \bibinfo {author} {\bibfnamefont {R.}~\bibnamefont {Wittkowski}},\ }\href@noop {} {\bibfield  {journal} {\bibinfo  {journal} {Physical Review Letters}\ }\textbf {\bibinfo {volume} {131}},\ \bibinfo {pages} {168203} (\bibinfo {year} {2023})}\BibitemShut {NoStop}%
\bibitem [{\citenamefont {Safferling}\ \emph {et~al.}(2013)\citenamefont {Safferling}, \citenamefont {S{\"u}tterlin}, \citenamefont {Westphal}, \citenamefont {Ernst}, \citenamefont {Breuhahn}, \citenamefont {James}, \citenamefont {J{\"a}ger}, \citenamefont {Halama},\ and\ \citenamefont {Grabe}}]{safferling2013wound}%
  \BibitemOpen
  \bibfield  {author} {\bibinfo {author} {\bibfnamefont {K.}~\bibnamefont {Safferling}}, \bibinfo {author} {\bibfnamefont {T.}~\bibnamefont {S{\"u}tterlin}}, \bibinfo {author} {\bibfnamefont {K.}~\bibnamefont {Westphal}}, \bibinfo {author} {\bibfnamefont {C.}~\bibnamefont {Ernst}}, \bibinfo {author} {\bibfnamefont {K.}~\bibnamefont {Breuhahn}}, \bibinfo {author} {\bibfnamefont {M.}~\bibnamefont {James}}, \bibinfo {author} {\bibfnamefont {D.}~\bibnamefont {J{\"a}ger}}, \bibinfo {author} {\bibfnamefont {N.}~\bibnamefont {Halama}}, \ and\ \bibinfo {author} {\bibfnamefont {N.}~\bibnamefont {Grabe}},\ }\href@noop {} {\bibfield  {journal} {\bibinfo  {journal} {Journal of Cell Biology}\ }\textbf {\bibinfo {volume} {203}},\ \bibinfo {pages} {691} (\bibinfo {year} {2013})}\BibitemShut {NoStop}%
\bibitem [{\citenamefont {Chen}\ and\ \citenamefont {Zhang}(2022)}]{PhysRevE.106.L052903}%
  \BibitemOpen
  \bibfield  {author} {\bibinfo {author} {\bibfnamefont {Y.}~\bibnamefont {Chen}}\ and\ \bibinfo {author} {\bibfnamefont {J.}~\bibnamefont {Zhang}},\ }\href {\doibase 10.1103/PhysRevE.106.L052903} {\bibfield  {journal} {\bibinfo  {journal} {Phys. Rev. E}\ }\textbf {\bibinfo {volume} {106}},\ \bibinfo {pages} {L052903} (\bibinfo {year} {2022})}\BibitemShut {NoStop}%
\bibitem [{\citenamefont {Chen}\ and\ \citenamefont {Zhang}(2024)}]{chen2023anomalous}%
  \BibitemOpen
  \bibfield  {author} {\bibinfo {author} {\bibfnamefont {Y.}~\bibnamefont {Chen}}\ and\ \bibinfo {author} {\bibfnamefont {J.}~\bibnamefont {Zhang}},\ }\href@noop {} {\bibfield  {journal} {\bibinfo  {journal} {Nature Communications}\ }\textbf {\bibinfo {volume} {15}},\ \bibinfo {pages} {6032} (\bibinfo {year} {2024})}\BibitemShut {NoStop}%
\bibitem [{\citenamefont {Kudrolli}\ \emph {et~al.}(2008)\citenamefont {Kudrolli}, \citenamefont {Lumay}, \citenamefont {Volfson},\ and\ \citenamefont {Tsimring}}]{kudrolli-PhysRevLett.100.058001}%
  \BibitemOpen
  \bibfield  {author} {\bibinfo {author} {\bibfnamefont {A.}~\bibnamefont {Kudrolli}}, \bibinfo {author} {\bibfnamefont {G.}~\bibnamefont {Lumay}}, \bibinfo {author} {\bibfnamefont {D.}~\bibnamefont {Volfson}}, \ and\ \bibinfo {author} {\bibfnamefont {L.~S.}\ \bibnamefont {Tsimring}},\ }\href {\doibase 10.1103/PhysRevLett.100.058001} {\bibfield  {journal} {\bibinfo  {journal} {Phys. Rev. Lett.}\ }\textbf {\bibinfo {volume} {100}},\ \bibinfo {pages} {058001} (\bibinfo {year} {2008})}\BibitemShut {NoStop}%
\bibitem [{\citenamefont {Narayan}\ \emph {et~al.}(2007)\citenamefont {Narayan}, \citenamefont {Ramaswamy},\ and\ \citenamefont {Menon}}]{narayan2007long}%
  \BibitemOpen
  \bibfield  {author} {\bibinfo {author} {\bibfnamefont {V.}~\bibnamefont {Narayan}}, \bibinfo {author} {\bibfnamefont {S.}~\bibnamefont {Ramaswamy}}, \ and\ \bibinfo {author} {\bibfnamefont {N.}~\bibnamefont {Menon}},\ }\href@noop {} {\bibfield  {journal} {\bibinfo  {journal} {Science}\ }\textbf {\bibinfo {volume} {317}},\ \bibinfo {pages} {105} (\bibinfo {year} {2007})}\BibitemShut {NoStop}%
\bibitem [{\citenamefont {Deseigne}\ \emph {et~al.}(2010)\citenamefont {Deseigne}, \citenamefont {Dauchot},\ and\ \citenamefont {Chat\'e}}]{dauchot-PhysRevLett.105.098001}%
  \BibitemOpen
  \bibfield  {author} {\bibinfo {author} {\bibfnamefont {J.}~\bibnamefont {Deseigne}}, \bibinfo {author} {\bibfnamefont {O.}~\bibnamefont {Dauchot}}, \ and\ \bibinfo {author} {\bibfnamefont {H.}~\bibnamefont {Chat\'e}},\ }\href {\doibase 10.1103/PhysRevLett.105.098001} {\bibfield  {journal} {\bibinfo  {journal} {Phys. Rev. Lett.}\ }\textbf {\bibinfo {volume} {105}},\ \bibinfo {pages} {098001} (\bibinfo {year} {2010})}\BibitemShut {NoStop}%
\bibitem [{\citenamefont {Scholz}\ and\ \citenamefont {P\"oschel}(2017)}]{scholz-PhysRevLett.118.198003}%
  \BibitemOpen
  \bibfield  {author} {\bibinfo {author} {\bibfnamefont {C.}~\bibnamefont {Scholz}}\ and\ \bibinfo {author} {\bibfnamefont {T.}~\bibnamefont {P\"oschel}},\ }\href {\doibase 10.1103/PhysRevLett.118.198003} {\bibfield  {journal} {\bibinfo  {journal} {Phys. Rev. Lett.}\ }\textbf {\bibinfo {volume} {118}},\ \bibinfo {pages} {198003} (\bibinfo {year} {2017})}\BibitemShut {NoStop}%
\bibitem [{\citenamefont {Jiang}\ \emph {et~al.}(2024{\natexlab{a}})\citenamefont {Jiang}, \citenamefont {Zheng}, \citenamefont {Chen}, \citenamefont {Baggioli},\ and\ \citenamefont {Zhang}}]{2403.08285}%
  \BibitemOpen
  \bibfield  {author} {\bibinfo {author} {\bibfnamefont {C.}~\bibnamefont {Jiang}}, \bibinfo {author} {\bibfnamefont {Z.}~\bibnamefont {Zheng}}, \bibinfo {author} {\bibfnamefont {Y.}~\bibnamefont {Chen}}, \bibinfo {author} {\bibfnamefont {M.}~\bibnamefont {Baggioli}}, \ and\ \bibinfo {author} {\bibfnamefont {J.}~\bibnamefont {Zhang}},\ }\href@noop {} {\enquote {\bibinfo {title} {Experimental observation of gapped shear waves and liquid-like to gas-like dynamical crossover in active granular matter},}\ } (\bibinfo {year} {2024}{\natexlab{a}}),\ \Eprint {http://arxiv.org/abs/arXiv:2403.08285} {arXiv:2403.08285} \BibitemShut {NoStop}%
\bibitem [{\citenamefont {Chaikin}\ and\ \citenamefont {Lubensky}(2000)}]{chaikin2000principles}%
  \BibitemOpen
  \bibfield  {author} {\bibinfo {author} {\bibfnamefont {P.}~\bibnamefont {Chaikin}}\ and\ \bibinfo {author} {\bibfnamefont {T.}~\bibnamefont {Lubensky}},\ }\href {https://books.google.gr/books?id=P9YjNjzr9OIC} {\emph {\bibinfo {title} {Principles of Condensed Matter Physics}}}\ (\bibinfo  {publisher} {Cambridge University Press},\ \bibinfo {year} {2000})\BibitemShut {NoStop}%
\bibitem [{\citenamefont {P.~w. Anderson}\ and\ \citenamefont {c.~M.~Varma}(1972)}]{doi:10.1080/14786437208229210}%
  \BibitemOpen
  \bibfield  {author} {\bibinfo {author} {\bibfnamefont {B.~I.~H.}\ \bibnamefont {P.~w. Anderson}}\ and\ \bibinfo {author} {\bibnamefont {c.~M.~Varma}},\ }\href {\doibase 10.1080/14786437208229210} {\bibfield  {journal} {\bibinfo  {journal} {The Philosophical Magazine: A Journal of Theoretical Experimental and Applied Physics}\ }\textbf {\bibinfo {volume} {25}},\ \bibinfo {pages} {1} (\bibinfo {year} {1972})}\BibitemShut {NoStop}%
\bibitem [{\citenamefont {Phillips}(1972)}]{Phillips1972}%
  \BibitemOpen
  \bibfield  {author} {\bibinfo {author} {\bibfnamefont {W.~A.}\ \bibnamefont {Phillips}},\ }\href {\doibase 10.1007/BF00660072} {\bibfield  {journal} {\bibinfo  {journal} {Journal of Low Temperature Physics}\ }\textbf {\bibinfo {volume} {7}},\ \bibinfo {pages} {351} (\bibinfo {year} {1972})}\BibitemShut {NoStop}%
\bibitem [{\citenamefont {Zeller}\ and\ \citenamefont {Pohl}(1971)}]{PhysRevB.4.2029}%
  \BibitemOpen
  \bibfield  {author} {\bibinfo {author} {\bibfnamefont {R.~C.}\ \bibnamefont {Zeller}}\ and\ \bibinfo {author} {\bibfnamefont {R.~O.}\ \bibnamefont {Pohl}},\ }\href {\doibase 10.1103/PhysRevB.4.2029} {\bibfield  {journal} {\bibinfo  {journal} {Phys. Rev. B}\ }\textbf {\bibinfo {volume} {4}},\ \bibinfo {pages} {2029} (\bibinfo {year} {1971})}\BibitemShut {NoStop}%
\bibitem [{\citenamefont {Lerner}\ and\ \citenamefont {Bouchbinder}(2021)}]{10.1063/5.0069477}%
  \BibitemOpen
  \bibfield  {author} {\bibinfo {author} {\bibfnamefont {E.}~\bibnamefont {Lerner}}\ and\ \bibinfo {author} {\bibfnamefont {E.}~\bibnamefont {Bouchbinder}},\ }\href {\doibase 10.1063/5.0069477} {\bibfield  {journal} {\bibinfo  {journal} {The Journal of Chemical Physics}\ }\textbf {\bibinfo {volume} {155}},\ \bibinfo {pages} {200901} (\bibinfo {year} {2021})}\BibitemShut {NoStop}%
\bibitem [{\citenamefont {Ramos}(2022)}]{ramos2022low}%
  \BibitemOpen
  \bibfield  {author} {\bibinfo {author} {\bibfnamefont {M.~A.}\ \bibnamefont {Ramos}},\ }\href {\doibase 10.1142/q0371} {\emph {\bibinfo {title} {Low-temperature Thermal and Vibrational Properties of Disordered Solids: A Half-century of Universal" anomalies" of Glasses}}}\ (\bibinfo  {publisher} {World Scientific},\ \bibinfo {year} {2022})\BibitemShut {NoStop}%
\bibitem [{\citenamefont {Argon}(1979)}]{ARGON197947}%
  \BibitemOpen
  \bibfield  {author} {\bibinfo {author} {\bibfnamefont {A.}~\bibnamefont {Argon}},\ }\href {\doibase https://doi.org/10.1016/0001-6160(79)90055-5} {\bibfield  {journal} {\bibinfo  {journal} {Acta Metallurgica}\ }\textbf {\bibinfo {volume} {27}},\ \bibinfo {pages} {47} (\bibinfo {year} {1979})}\BibitemShut {NoStop}%
\bibitem [{\citenamefont {Eshelby}\ and\ \citenamefont {Peierls}(1959)}]{doi:10.1098/rspa.1959.0173}%
  \BibitemOpen
  \bibfield  {author} {\bibinfo {author} {\bibfnamefont {J.~D.}\ \bibnamefont {Eshelby}}\ and\ \bibinfo {author} {\bibfnamefont {R.~E.}\ \bibnamefont {Peierls}},\ }\href {\doibase 10.1098/rspa.1959.0173} {\bibfield  {journal} {\bibinfo  {journal} {Proceedings of the Royal Society of London. Series A. Mathematical and Physical Sciences}\ }\textbf {\bibinfo {volume} {252}},\ \bibinfo {pages} {561} (\bibinfo {year} {1959})}\BibitemShut {NoStop}%
\bibitem [{\citenamefont {Baggioli}(2023)}]{Baggioli2023}%
  \BibitemOpen
  \bibfield  {author} {\bibinfo {author} {\bibfnamefont {M.}~\bibnamefont {Baggioli}},\ }\href {\doibase 10.1038/s41467-023-38549-8} {\bibfield  {journal} {\bibinfo  {journal} {Nature Communications}\ }\textbf {\bibinfo {volume} {14}},\ \bibinfo {pages} {2956} (\bibinfo {year} {2023})}\BibitemShut {NoStop}%
\bibitem [{\citenamefont {Falk}\ and\ \citenamefont {Langer}(2011)}]{annurev:/content/journals/10.1146/annurev-conmatphys-062910-140452}%
  \BibitemOpen
  \bibfield  {author} {\bibinfo {author} {\bibfnamefont {M.~L.}\ \bibnamefont {Falk}}\ and\ \bibinfo {author} {\bibfnamefont {J.}~\bibnamefont {Langer}},\ }\href {\doibase https://doi.org/10.1146/annurev-conmatphys-062910-140452} {\bibfield  {journal} {\bibinfo  {journal} {Annual Review of Condensed Matter Physics}\ }\textbf {\bibinfo {volume} {2}},\ \bibinfo {pages} {353} (\bibinfo {year} {2011})}\BibitemShut {NoStop}%
\bibitem [{\citenamefont {Kleinert}(1989)}]{kleinert1989gauge}%
  \BibitemOpen
  \bibfield  {author} {\bibinfo {author} {\bibfnamefont {H.}~\bibnamefont {Kleinert}},\ }\href@noop {} {\emph {\bibinfo {title} {Gauge Fields in Condensed Matter: Vol. 1: Superflow and Vortex Lines (Disorder Fields, Phase Transitions) Vol. 2: Stresses and Defects (Differential Geometry, Crystal Melting)}}}\ (\bibinfo  {publisher} {World Scientific},\ \bibinfo {year} {1989})\BibitemShut {NoStop}%
\bibitem [{\citenamefont {Donati}\ \emph {et~al.}(1998)\citenamefont {Donati}, \citenamefont {Douglas}, \citenamefont {Kob}, \citenamefont {Plimpton}, \citenamefont {Poole},\ and\ \citenamefont {Glotzer}}]{PhysRevLett.80.2338}%
  \BibitemOpen
  \bibfield  {author} {\bibinfo {author} {\bibfnamefont {C.}~\bibnamefont {Donati}}, \bibinfo {author} {\bibfnamefont {J.~F.}\ \bibnamefont {Douglas}}, \bibinfo {author} {\bibfnamefont {W.}~\bibnamefont {Kob}}, \bibinfo {author} {\bibfnamefont {S.~J.}\ \bibnamefont {Plimpton}}, \bibinfo {author} {\bibfnamefont {P.~H.}\ \bibnamefont {Poole}}, \ and\ \bibinfo {author} {\bibfnamefont {S.~C.}\ \bibnamefont {Glotzer}},\ }\href {\doibase 10.1103/PhysRevLett.80.2338} {\bibfield  {journal} {\bibinfo  {journal} {Phys. Rev. Lett.}\ }\textbf {\bibinfo {volume} {80}},\ \bibinfo {pages} {2338} (\bibinfo {year} {1998})}\BibitemShut {NoStop}%
\bibitem [{\citenamefont {Pazmino~Betancourt}\ \emph {et~al.}(2014)\citenamefont {Pazmino~Betancourt}, \citenamefont {Douglas},\ and\ \citenamefont {Starr}}]{10.1063/1.4878502}%
  \BibitemOpen
  \bibfield  {author} {\bibinfo {author} {\bibfnamefont {B.~A.}\ \bibnamefont {Pazmino~Betancourt}}, \bibinfo {author} {\bibfnamefont {J.~F.}\ \bibnamefont {Douglas}}, \ and\ \bibinfo {author} {\bibfnamefont {F.~W.}\ \bibnamefont {Starr}},\ }\href {\doibase 10.1063/1.4878502} {\bibfield  {journal} {\bibinfo  {journal} {The Journal of Chemical Physics}\ }\textbf {\bibinfo {volume} {140}},\ \bibinfo {pages} {204509} (\bibinfo {year} {2014})}\BibitemShut {NoStop}%
\bibitem [{\citenamefont {Zhang}\ \emph {et~al.}(2021{\natexlab{a}})\citenamefont {Zhang}, \citenamefont {Wang}, \citenamefont {Yu},\ and\ \citenamefont {Douglas}}]{10.1063/5.0039162}%
  \BibitemOpen
  \bibfield  {author} {\bibinfo {author} {\bibfnamefont {H.}~\bibnamefont {Zhang}}, \bibinfo {author} {\bibfnamefont {X.}~\bibnamefont {Wang}}, \bibinfo {author} {\bibfnamefont {H.-B.}\ \bibnamefont {Yu}}, \ and\ \bibinfo {author} {\bibfnamefont {J.~F.}\ \bibnamefont {Douglas}},\ }\href {\doibase 10.1063/5.0039162} {\bibfield  {journal} {\bibinfo  {journal} {Journal of Chemical Physics}\ }\textbf {\bibinfo {volume} {154}},\ \bibinfo {pages} {084505} (\bibinfo {year} {2021}{\natexlab{a}})}\BibitemShut {NoStop}%
\bibitem [{\citenamefont {Zhang}\ \emph {et~al.}(2015)\citenamefont {Zhang}, \citenamefont {Zhong}, \citenamefont {Douglas}, \citenamefont {Wang}, \citenamefont {Cao}, \citenamefont {Zhang},\ and\ \citenamefont {Jiang}}]{10.1063/1.4918807}%
  \BibitemOpen
  \bibfield  {author} {\bibinfo {author} {\bibfnamefont {H.}~\bibnamefont {Zhang}}, \bibinfo {author} {\bibfnamefont {C.}~\bibnamefont {Zhong}}, \bibinfo {author} {\bibfnamefont {J.~F.}\ \bibnamefont {Douglas}}, \bibinfo {author} {\bibfnamefont {X.}~\bibnamefont {Wang}}, \bibinfo {author} {\bibfnamefont {Q.}~\bibnamefont {Cao}}, \bibinfo {author} {\bibfnamefont {D.}~\bibnamefont {Zhang}}, \ and\ \bibinfo {author} {\bibfnamefont {J.-Z.}\ \bibnamefont {Jiang}},\ }\href {\doibase 10.1063/1.4918807} {\bibfield  {journal} {\bibinfo  {journal} {The Journal of Chemical Physics}\ }\textbf {\bibinfo {volume} {142}},\ \bibinfo {pages} {164506} (\bibinfo {year} {2015})}\BibitemShut {NoStop}%
\bibitem [{\citenamefont {Zhang}\ \emph {et~al.}(2013)\citenamefont {Zhang}, \citenamefont {Khalkhali}, \citenamefont {Liu},\ and\ \citenamefont {Douglas}}]{10.1063/1.4769267}%
  \BibitemOpen
  \bibfield  {author} {\bibinfo {author} {\bibfnamefont {H.}~\bibnamefont {Zhang}}, \bibinfo {author} {\bibfnamefont {M.}~\bibnamefont {Khalkhali}}, \bibinfo {author} {\bibfnamefont {Q.}~\bibnamefont {Liu}}, \ and\ \bibinfo {author} {\bibfnamefont {J.~F.}\ \bibnamefont {Douglas}},\ }\href {\doibase 10.1063/1.4769267} {\bibfield  {journal} {\bibinfo  {journal} {The Journal of Chemical Physics}\ }\textbf {\bibinfo {volume} {138}},\ \bibinfo {pages} {12A538} (\bibinfo {year} {2013})}\BibitemShut {NoStop}%
\bibitem [{\citenamefont {Zhao}\ \emph {et~al.}(2024)\citenamefont {Zhao}, \citenamefont {Baggioli}, \citenamefont {Xu}, \citenamefont {Douglas},\ and\ \citenamefont {Wang}}]{zhao2024quasicrystals}%
  \BibitemOpen
  \bibfield  {author} {\bibinfo {author} {\bibfnamefont {K.}~\bibnamefont {Zhao}}, \bibinfo {author} {\bibfnamefont {M.}~\bibnamefont {Baggioli}}, \bibinfo {author} {\bibfnamefont {W.-S.}\ \bibnamefont {Xu}}, \bibinfo {author} {\bibfnamefont {J.~F.}\ \bibnamefont {Douglas}}, \ and\ \bibinfo {author} {\bibfnamefont {Y.-J.}\ \bibnamefont {Wang}},\ }\href@noop {} {\bibfield  {journal} {\bibinfo  {journal} {arXiv preprint arXiv:2402.10295}\ } (\bibinfo {year} {2024})}\BibitemShut {NoStop}%
\bibitem [{\citenamefont {Hung}\ and\ \citenamefont {Simmons}(2020)}]{doi:10.1021/acs.jpcb.9b09468}%
  \BibitemOpen
  \bibfield  {author} {\bibinfo {author} {\bibfnamefont {J.-H.}\ \bibnamefont {Hung}}\ and\ \bibinfo {author} {\bibfnamefont {D.~S.}\ \bibnamefont {Simmons}},\ }\href {\doibase 10.1021/acs.jpcb.9b09468} {\bibfield  {journal} {\bibinfo  {journal} {The Journal of Physical Chemistry B}\ }\textbf {\bibinfo {volume} {124}},\ \bibinfo {pages} {266} (\bibinfo {year} {2020})},\ \bibinfo {note} {pMID: 31886663}\BibitemShut {NoStop}%
\bibitem [{\citenamefont {Bianchi}\ \emph {et~al.}(2020)\citenamefont {Bianchi}, \citenamefont {Giordano},\ and\ \citenamefont {Lund}}]{PhysRevB.101.174311}%
  \BibitemOpen
  \bibfield  {author} {\bibinfo {author} {\bibfnamefont {E.}~\bibnamefont {Bianchi}}, \bibinfo {author} {\bibfnamefont {V.~M.}\ \bibnamefont {Giordano}}, \ and\ \bibinfo {author} {\bibfnamefont {F.}~\bibnamefont {Lund}},\ }\href {\doibase 10.1103/PhysRevB.101.174311} {\bibfield  {journal} {\bibinfo  {journal} {Phys. Rev. B}\ }\textbf {\bibinfo {volume} {101}},\ \bibinfo {pages} {174311} (\bibinfo {year} {2020})}\BibitemShut {NoStop}%
\bibitem [{\citenamefont {Hu}\ and\ \citenamefont {Tanaka}(2022)}]{Hu2022}%
  \BibitemOpen
  \bibfield  {author} {\bibinfo {author} {\bibfnamefont {Y.-C.}\ \bibnamefont {Hu}}\ and\ \bibinfo {author} {\bibfnamefont {H.}~\bibnamefont {Tanaka}},\ }\href {\doibase 10.1038/s41567-022-01628-6} {\bibfield  {journal} {\bibinfo  {journal} {Nature Physics}\ }\textbf {\bibinfo {volume} {18}},\ \bibinfo {pages} {669} (\bibinfo {year} {2022})}\BibitemShut {NoStop}%
\bibitem [{\citenamefont {Hu}\ and\ \citenamefont {Tanaka}(2023)}]{PhysRevResearch.5.023055}%
  \BibitemOpen
  \bibfield  {author} {\bibinfo {author} {\bibfnamefont {Y.-C.}\ \bibnamefont {Hu}}\ and\ \bibinfo {author} {\bibfnamefont {H.}~\bibnamefont {Tanaka}},\ }\href {\doibase 10.1103/PhysRevResearch.5.023055} {\bibfield  {journal} {\bibinfo  {journal} {Phys. Rev. Res.}\ }\textbf {\bibinfo {volume} {5}},\ \bibinfo {pages} {023055} (\bibinfo {year} {2023})}\BibitemShut {NoStop}%
\bibitem [{\citenamefont {Zhang}\ \emph {et~al.}(2021{\natexlab{b}})\citenamefont {Zhang}, \citenamefont {Wang}, \citenamefont {Yu},\ and\ \citenamefont {Douglas}}]{d1}%
  \BibitemOpen
  \bibfield  {author} {\bibinfo {author} {\bibfnamefont {H.}~\bibnamefont {Zhang}}, \bibinfo {author} {\bibfnamefont {X.}~\bibnamefont {Wang}}, \bibinfo {author} {\bibfnamefont {H.-B.}\ \bibnamefont {Yu}}, \ and\ \bibinfo {author} {\bibfnamefont {J.~F.}\ \bibnamefont {Douglas}},\ }\href {\doibase 10.1063/5.0039162} {\bibfield  {journal} {\bibinfo  {journal} {The Journal of Chemical Physics}\ }\textbf {\bibinfo {volume} {154}},\ \bibinfo {pages} {084505} (\bibinfo {year} {2021}{\natexlab{b}})}\BibitemShut {NoStop}%
\bibitem [{\citenamefont {Jiang}\ \emph {et~al.}(2024{\natexlab{b}})\citenamefont {Jiang}, \citenamefont {Baggioli},\ and\ \citenamefont {Douglas}}]{jiang2023stringlet}%
  \BibitemOpen
  \bibfield  {author} {\bibinfo {author} {\bibfnamefont {C.}~\bibnamefont {Jiang}}, \bibinfo {author} {\bibfnamefont {M.}~\bibnamefont {Baggioli}}, \ and\ \bibinfo {author} {\bibfnamefont {J.~F.}\ \bibnamefont {Douglas}},\ }\href {\doibase 10.1063/5.0210057} {\bibfield  {journal} {\bibinfo  {journal} {The Journal of Chemical Physics}\ }\textbf {\bibinfo {volume} {160}},\ \bibinfo {pages} {214505} (\bibinfo {year} {2024}{\natexlab{b}})}\BibitemShut {NoStop}%
\bibitem [{\citenamefont {Jiang}\ and\ \citenamefont {Baggioli}(2024)}]{jiang2024phonons}%
  \BibitemOpen
  \bibfield  {author} {\bibinfo {author} {\bibfnamefont {C.}~\bibnamefont {Jiang}}\ and\ \bibinfo {author} {\bibfnamefont {M.}~\bibnamefont {Baggioli}},\ }\href@noop {} {\bibfield  {journal} {\bibinfo  {journal} {arXiv preprint arXiv:2403.18221}\ } (\bibinfo {year} {2024})}\BibitemShut {NoStop}%
\bibitem [{\citenamefont {Zhang}\ \emph {et~al.}(2024)\citenamefont {Zhang}, \citenamefont {Zhang},\ and\ \citenamefont {Douglas}}]{10.1063/5.0197386}%
  \BibitemOpen
  \bibfield  {author} {\bibinfo {author} {\bibfnamefont {J.}~\bibnamefont {Zhang}}, \bibinfo {author} {\bibfnamefont {H.}~\bibnamefont {Zhang}}, \ and\ \bibinfo {author} {\bibfnamefont {J.~F.}\ \bibnamefont {Douglas}},\ }\href {\doibase 10.1063/5.0197386} {\bibfield  {journal} {\bibinfo  {journal} {The Journal of Chemical Physics}\ }\textbf {\bibinfo {volume} {160}},\ \bibinfo {pages} {114506} (\bibinfo {year} {2024})}\BibitemShut {NoStop}%
\bibitem [{\citenamefont {T{\o}mterud}\ \emph {et~al.}(2023)\citenamefont {T{\o}mterud}, \citenamefont {Eder}, \citenamefont {B{\"u}chner}, \citenamefont {Wondraczek}, \citenamefont {Simonsen}, \citenamefont {Schirmacher}, \citenamefont {Manson},\ and\ \citenamefont {Holst}}]{Tomterud2023}%
  \BibitemOpen
  \bibfield  {author} {\bibinfo {author} {\bibfnamefont {M.}~\bibnamefont {T{\o}mterud}}, \bibinfo {author} {\bibfnamefont {S.~D.}\ \bibnamefont {Eder}}, \bibinfo {author} {\bibfnamefont {C.}~\bibnamefont {B{\"u}chner}}, \bibinfo {author} {\bibfnamefont {L.}~\bibnamefont {Wondraczek}}, \bibinfo {author} {\bibfnamefont {I.}~\bibnamefont {Simonsen}}, \bibinfo {author} {\bibfnamefont {W.}~\bibnamefont {Schirmacher}}, \bibinfo {author} {\bibfnamefont {J.~R.}\ \bibnamefont {Manson}}, \ and\ \bibinfo {author} {\bibfnamefont {B.}~\bibnamefont {Holst}},\ }\href {\doibase 10.1038/s41567-023-02177-2} {\bibfield  {journal} {\bibinfo  {journal} {Nature Physics}\ }\textbf {\bibinfo {volume} {19}},\ \bibinfo {pages} {1910} (\bibinfo {year} {2023})}\BibitemShut {NoStop}%
\bibitem [{\citenamefont {Berardi}\ \emph {et~al.}(2010)\citenamefont {Berardi}, \citenamefont {Barros}, \citenamefont {Douglas},\ and\ \citenamefont {Losert}}]{PhysRevE.81.041301}%
  \BibitemOpen
  \bibfield  {author} {\bibinfo {author} {\bibfnamefont {C.~R.}\ \bibnamefont {Berardi}}, \bibinfo {author} {\bibfnamefont {K.}~\bibnamefont {Barros}}, \bibinfo {author} {\bibfnamefont {J.~F.}\ \bibnamefont {Douglas}}, \ and\ \bibinfo {author} {\bibfnamefont {W.}~\bibnamefont {Losert}},\ }\href {\doibase 10.1103/PhysRevE.81.041301} {\bibfield  {journal} {\bibinfo  {journal} {Phys. Rev. E}\ }\textbf {\bibinfo {volume} {81}},\ \bibinfo {pages} {041301} (\bibinfo {year} {2010})}\BibitemShut {NoStop}%
\bibitem [{\citenamefont {Zhang}\ \emph {et~al.}(2009)\citenamefont {Zhang}, \citenamefont {Srolovitz}, \citenamefont {Douglas},\ and\ \citenamefont {Warren}}]{doi:10.1073/pnas.0900227106}%
  \BibitemOpen
  \bibfield  {author} {\bibinfo {author} {\bibfnamefont {H.}~\bibnamefont {Zhang}}, \bibinfo {author} {\bibfnamefont {D.~J.}\ \bibnamefont {Srolovitz}}, \bibinfo {author} {\bibfnamefont {J.~F.}\ \bibnamefont {Douglas}}, \ and\ \bibinfo {author} {\bibfnamefont {J.~A.}\ \bibnamefont {Warren}},\ }\href {\doibase 10.1073/pnas.0900227106} {\bibfield  {journal} {\bibinfo  {journal} {Proceedings of the National Academy of Sciences}\ }\textbf {\bibinfo {volume} {106}},\ \bibinfo {pages} {7735} (\bibinfo {year} {2009})}\BibitemShut {NoStop}%
\bibitem [{\citenamefont {Nagamanasa}\ \emph {et~al.}(2011)\citenamefont {Nagamanasa}, \citenamefont {Gokhale}, \citenamefont {Ganapathy},\ and\ \citenamefont {Sood}}]{doi:10.1073/pnas.1101858108}%
  \BibitemOpen
  \bibfield  {author} {\bibinfo {author} {\bibfnamefont {K.~H.}\ \bibnamefont {Nagamanasa}}, \bibinfo {author} {\bibfnamefont {S.}~\bibnamefont {Gokhale}}, \bibinfo {author} {\bibfnamefont {R.}~\bibnamefont {Ganapathy}}, \ and\ \bibinfo {author} {\bibfnamefont {A.~K.}\ \bibnamefont {Sood}},\ }\href {\doibase 10.1073/pnas.1101858108} {\bibfield  {journal} {\bibinfo  {journal} {Proceedings of the National Academy of Sciences}\ }\textbf {\bibinfo {volume} {108}},\ \bibinfo {pages} {11323} (\bibinfo {year} {2011})}\BibitemShut {NoStop}%
\bibitem [{\citenamefont {Ramaswamy}(2010)}]{ramaswamy2010mechanics}%
  \BibitemOpen
  \bibfield  {author} {\bibinfo {author} {\bibfnamefont {S.}~\bibnamefont {Ramaswamy}},\ }\href@noop {} {\bibfield  {journal} {\bibinfo  {journal} {Annu. Rev. Condens. Matter Phys.}\ }\textbf {\bibinfo {volume} {1}},\ \bibinfo {pages} {323} (\bibinfo {year} {2010})}\BibitemShut {NoStop}%
\bibitem [{\citenamefont {Binder}\ and\ \citenamefont {Kob}(2011)}]{binder2011glassy}%
  \BibitemOpen
  \bibfield  {author} {\bibinfo {author} {\bibfnamefont {K.}~\bibnamefont {Binder}}\ and\ \bibinfo {author} {\bibfnamefont {W.}~\bibnamefont {Kob}},\ }\href@noop {} {\emph {\bibinfo {title} {Glassy materials and disordered solids: An introduction to their statistical mechanics}}}\ (\bibinfo  {publisher} {World scientific},\ \bibinfo {year} {2011})\BibitemShut {NoStop}%
\bibitem [{\citenamefont {Malinovsky}\ \emph {et~al.}(1991)\citenamefont {Malinovsky}, \citenamefont {Novikov},\ and\ \citenamefont {Sokolov}}]{MALINOVSKY199163}%
  \BibitemOpen
  \bibfield  {author} {\bibinfo {author} {\bibfnamefont {V.}~\bibnamefont {Malinovsky}}, \bibinfo {author} {\bibfnamefont {V.}~\bibnamefont {Novikov}}, \ and\ \bibinfo {author} {\bibfnamefont {A.}~\bibnamefont {Sokolov}},\ }\href {\doibase https://doi.org/10.1016/0375-9601(91)90363-D} {\bibfield  {journal} {\bibinfo  {journal} {Physics Letters A}\ }\textbf {\bibinfo {volume} {153}},\ \bibinfo {pages} {63} (\bibinfo {year} {1991})}\BibitemShut {NoStop}%
\bibitem [{\citenamefont {Baggioli}\ \emph {et~al.}(2020)\citenamefont {Baggioli}, \citenamefont {Vasin}, \citenamefont {Brazhkin},\ and\ \citenamefont {Trachenko}}]{BAGGIOLI20201}%
  \BibitemOpen
  \bibfield  {author} {\bibinfo {author} {\bibfnamefont {M.}~\bibnamefont {Baggioli}}, \bibinfo {author} {\bibfnamefont {M.}~\bibnamefont {Vasin}}, \bibinfo {author} {\bibfnamefont {V.}~\bibnamefont {Brazhkin}}, \ and\ \bibinfo {author} {\bibfnamefont {K.}~\bibnamefont {Trachenko}},\ }\href {\doibase https://doi.org/10.1016/j.physrep.2020.04.002} {\bibfield  {journal} {\bibinfo  {journal} {Physics Reports}\ }\textbf {\bibinfo {volume} {865}},\ \bibinfo {pages} {1} (\bibinfo {year} {2020})},\ \bibinfo {note} {gapped momentum states}\BibitemShut {NoStop}%
\bibitem [{\citenamefont {Chumakov}\ \emph {et~al.}(2004)\citenamefont {Chumakov}, \citenamefont {Sergueev}, \citenamefont {van B\"urck}, \citenamefont {Schirmacher}, \citenamefont {Asthalter}, \citenamefont {R\"uffer}, \citenamefont {Leupold},\ and\ \citenamefont {Petry}}]{PhysRevLett.92.245508}%
  \BibitemOpen
  \bibfield  {author} {\bibinfo {author} {\bibfnamefont {A.~I.}\ \bibnamefont {Chumakov}}, \bibinfo {author} {\bibfnamefont {I.}~\bibnamefont {Sergueev}}, \bibinfo {author} {\bibfnamefont {U.}~\bibnamefont {van B\"urck}}, \bibinfo {author} {\bibfnamefont {W.}~\bibnamefont {Schirmacher}}, \bibinfo {author} {\bibfnamefont {T.}~\bibnamefont {Asthalter}}, \bibinfo {author} {\bibfnamefont {R.}~\bibnamefont {R\"uffer}}, \bibinfo {author} {\bibfnamefont {O.}~\bibnamefont {Leupold}}, \ and\ \bibinfo {author} {\bibfnamefont {W.}~\bibnamefont {Petry}},\ }\href {\doibase 10.1103/PhysRevLett.92.245508} {\bibfield  {journal} {\bibinfo  {journal} {Phys. Rev. Lett.}\ }\textbf {\bibinfo {volume} {92}},\ \bibinfo {pages} {245508} (\bibinfo {year} {2004})}\BibitemShut {NoStop}%
\bibitem [{\citenamefont {Kalampounias}\ \emph {et~al.}(2006)\citenamefont {Kalampounias}, \citenamefont {Yannopoulos},\ and\ \citenamefont {Papatheodorou}}]{10.1063/1.2360275}%
  \BibitemOpen
  \bibfield  {author} {\bibinfo {author} {\bibfnamefont {A.~G.}\ \bibnamefont {Kalampounias}}, \bibinfo {author} {\bibfnamefont {S.~N.}\ \bibnamefont {Yannopoulos}}, \ and\ \bibinfo {author} {\bibfnamefont {G.~N.}\ \bibnamefont {Papatheodorou}},\ }\href {\doibase 10.1063/1.2360275} {\bibfield  {journal} {\bibinfo  {journal} {The Journal of Chemical Physics}\ }\textbf {\bibinfo {volume} {125}},\ \bibinfo {pages} {164502} (\bibinfo {year} {2006})}\BibitemShut {NoStop}%
\bibitem [{\citenamefont {Hong}\ \emph {et~al.}(2011{\natexlab{a}})\citenamefont {Hong}, \citenamefont {Novikov},\ and\ \citenamefont {Sokolov}}]{PhysRevE.83.061508}%
  \BibitemOpen
  \bibfield  {author} {\bibinfo {author} {\bibfnamefont {L.}~\bibnamefont {Hong}}, \bibinfo {author} {\bibfnamefont {V.~N.}\ \bibnamefont {Novikov}}, \ and\ \bibinfo {author} {\bibfnamefont {A.~P.}\ \bibnamefont {Sokolov}},\ }\href {\doibase 10.1103/PhysRevE.83.061508} {\bibfield  {journal} {\bibinfo  {journal} {Phys. Rev. E}\ }\textbf {\bibinfo {volume} {83}},\ \bibinfo {pages} {061508} (\bibinfo {year} {2011}{\natexlab{a}})}\BibitemShut {NoStop}%
\bibitem [{\citenamefont {Hong}\ \emph {et~al.}(2011{\natexlab{b}})\citenamefont {Hong}, \citenamefont {Novikov},\ and\ \citenamefont {Sokolov}}]{HONG2011351}%
  \BibitemOpen
  \bibfield  {author} {\bibinfo {author} {\bibfnamefont {L.}~\bibnamefont {Hong}}, \bibinfo {author} {\bibfnamefont {V.}~\bibnamefont {Novikov}}, \ and\ \bibinfo {author} {\bibfnamefont {A.}~\bibnamefont {Sokolov}},\ }\href@noop {} {\bibfield  {journal} {\bibinfo  {journal} {Journal of Non-Crystalline Solids}\ }\textbf {\bibinfo {volume} {357}},\ \bibinfo {pages} {351} (\bibinfo {year} {2011}{\natexlab{b}})}\BibitemShut {NoStop}%
\bibitem [{\citenamefont {Malinovsky}\ \emph {et~al.}(1988)\citenamefont {Malinovsky}, \citenamefont {Novikov}, \citenamefont {Sokolov},\ and\ \citenamefont {Bagryansky}}]{MALINOVSKY1988111}%
  \BibitemOpen
  \bibfield  {author} {\bibinfo {author} {\bibfnamefont {V.}~\bibnamefont {Malinovsky}}, \bibinfo {author} {\bibfnamefont {V.}~\bibnamefont {Novikov}}, \bibinfo {author} {\bibfnamefont {A.}~\bibnamefont {Sokolov}}, \ and\ \bibinfo {author} {\bibfnamefont {V.}~\bibnamefont {Bagryansky}},\ }\href@noop {} {\bibfield  {journal} {\bibinfo  {journal} {Chemical Physics Letters}\ }\textbf {\bibinfo {volume} {143}},\ \bibinfo {pages} {111} (\bibinfo {year} {1988})}\BibitemShut {NoStop}%
\bibitem [{\citenamefont {Malinovsky}\ and\ \citenamefont {Sokolov}(1986)}]{MALINOVSKY1986757}%
  \BibitemOpen
  \bibfield  {author} {\bibinfo {author} {\bibfnamefont {V.}~\bibnamefont {Malinovsky}}\ and\ \bibinfo {author} {\bibfnamefont {A.}~\bibnamefont {Sokolov}},\ }\href@noop {} {\bibfield  {journal} {\bibinfo  {journal} {Solid State Communications}\ }\textbf {\bibinfo {volume} {57}},\ \bibinfo {pages} {757} (\bibinfo {year} {1986})}\BibitemShut {NoStop}%
\bibitem [{\citenamefont {Martin}\ and\ \citenamefont {Brenig}(1974)}]{https://doi.org/10.1002/pssb.2220640120}%
  \BibitemOpen
  \bibfield  {author} {\bibinfo {author} {\bibfnamefont {A.~J.}\ \bibnamefont {Martin}}\ and\ \bibinfo {author} {\bibfnamefont {W.}~\bibnamefont {Brenig}},\ }\href {\doibase https://doi.org/10.1002/pssb.2220640120} {\bibfield  {journal} {\bibinfo  {journal} {physica status solidi (b)}\ }\textbf {\bibinfo {volume} {64}},\ \bibinfo {pages} {163} (\bibinfo {year} {1974})}\BibitemShut {NoStop}%
\bibitem [{\citenamefont {Schirmacher}\ \emph {et~al.}(2008)\citenamefont {Schirmacher}, \citenamefont {Schmid}, \citenamefont {Tomaras}, \citenamefont {Viliani}, \citenamefont {Baldi}, \citenamefont {Ruocco},\ and\ \citenamefont {Scopigno}}]{https://doi.org/10.1002/pssc.200777584}%
  \BibitemOpen
  \bibfield  {author} {\bibinfo {author} {\bibfnamefont {W.}~\bibnamefont {Schirmacher}}, \bibinfo {author} {\bibfnamefont {B.}~\bibnamefont {Schmid}}, \bibinfo {author} {\bibfnamefont {C.}~\bibnamefont {Tomaras}}, \bibinfo {author} {\bibfnamefont {G.}~\bibnamefont {Viliani}}, \bibinfo {author} {\bibfnamefont {G.}~\bibnamefont {Baldi}}, \bibinfo {author} {\bibfnamefont {G.}~\bibnamefont {Ruocco}}, \ and\ \bibinfo {author} {\bibfnamefont {T.}~\bibnamefont {Scopigno}},\ }\href {\doibase https://doi.org/10.1002/pssc.200777584} {\bibfield  {journal} {\bibinfo  {journal} {physica status solidi c}\ }\textbf {\bibinfo {volume} {5}},\ \bibinfo {pages} {862} (\bibinfo {year} {2008})}\BibitemShut {NoStop}%
\bibitem [{\citenamefont {Duval}\ \emph {et~al.}(1990)\citenamefont {Duval}, \citenamefont {Boukenter},\ and\ \citenamefont {Achibat}}]{Duval_1990}%
  \BibitemOpen
  \bibfield  {author} {\bibinfo {author} {\bibfnamefont {E.}~\bibnamefont {Duval}}, \bibinfo {author} {\bibfnamefont {A.}~\bibnamefont {Boukenter}}, \ and\ \bibinfo {author} {\bibfnamefont {T.}~\bibnamefont {Achibat}},\ }\href {\doibase 10.1088/0953-8984/2/51/001} {\bibfield  {journal} {\bibinfo  {journal} {Journal of Physics: Condensed Matter}\ }\textbf {\bibinfo {volume} {2}},\ \bibinfo {pages} {10227} (\bibinfo {year} {1990})}\BibitemShut {NoStop}%
\bibitem [{\citenamefont {Sokolov}\ \emph {et~al.}(1992)\citenamefont {Sokolov}, \citenamefont {Kisliuk}, \citenamefont {Soltwisch},\ and\ \citenamefont {Quitmann}}]{PhysRevLett.69.1540}%
  \BibitemOpen
  \bibfield  {author} {\bibinfo {author} {\bibfnamefont {A.~P.}\ \bibnamefont {Sokolov}}, \bibinfo {author} {\bibfnamefont {A.}~\bibnamefont {Kisliuk}}, \bibinfo {author} {\bibfnamefont {M.}~\bibnamefont {Soltwisch}}, \ and\ \bibinfo {author} {\bibfnamefont {D.}~\bibnamefont {Quitmann}},\ }\href {\doibase 10.1103/PhysRevLett.69.1540} {\bibfield  {journal} {\bibinfo  {journal} {Phys. Rev. Lett.}\ }\textbf {\bibinfo {volume} {69}},\ \bibinfo {pages} {1540} (\bibinfo {year} {1992})}\BibitemShut {NoStop}%
\bibitem [{\citenamefont {Elliott}(1992)}]{Elliott_1992}%
  \BibitemOpen
  \bibfield  {author} {\bibinfo {author} {\bibfnamefont {S.~R.}\ \bibnamefont {Elliott}},\ }\href {\doibase 10.1209/0295-5075/19/3/009} {\bibfield  {journal} {\bibinfo  {journal} {Europhysics Letters}\ }\textbf {\bibinfo {volume} {19}},\ \bibinfo {pages} {201} (\bibinfo {year} {1992})}\BibitemShut {NoStop}%
\bibitem [{\citenamefont {Schmid}\ and\ \citenamefont {Schirmacher}(2008)}]{PhysRevLett.100.137402}%
  \BibitemOpen
  \bibfield  {author} {\bibinfo {author} {\bibfnamefont {B.}~\bibnamefont {Schmid}}\ and\ \bibinfo {author} {\bibfnamefont {W.}~\bibnamefont {Schirmacher}},\ }\href {\doibase 10.1103/PhysRevLett.100.137402} {\bibfield  {journal} {\bibinfo  {journal} {Phys. Rev. Lett.}\ }\textbf {\bibinfo {volume} {100}},\ \bibinfo {pages} {137402} (\bibinfo {year} {2008})}\BibitemShut {NoStop}%
\bibitem [{\citenamefont {Tanguy}\ \emph {et~al.}(2002)\citenamefont {Tanguy}, \citenamefont {Wittmer}, \citenamefont {Leonforte},\ and\ \citenamefont {Barrat}}]{PhysRevB.66.174205}%
  \BibitemOpen
  \bibfield  {author} {\bibinfo {author} {\bibfnamefont {A.}~\bibnamefont {Tanguy}}, \bibinfo {author} {\bibfnamefont {J.~P.}\ \bibnamefont {Wittmer}}, \bibinfo {author} {\bibfnamefont {F.}~\bibnamefont {Leonforte}}, \ and\ \bibinfo {author} {\bibfnamefont {J.-L.}\ \bibnamefont {Barrat}},\ }\href {\doibase 10.1103/PhysRevB.66.174205} {\bibfield  {journal} {\bibinfo  {journal} {Phys. Rev. B}\ }\textbf {\bibinfo {volume} {66}},\ \bibinfo {pages} {174205} (\bibinfo {year} {2002})}\BibitemShut {NoStop}%
\bibitem [{\citenamefont {Ghosh}\ \emph {et~al.}(2010)\citenamefont {Ghosh}, \citenamefont {Chikkadi}, \citenamefont {Schall}, \citenamefont {Kurchan},\ and\ \citenamefont {Bonn}}]{PhysRevLett.104.248305}%
  \BibitemOpen
  \bibfield  {author} {\bibinfo {author} {\bibfnamefont {A.}~\bibnamefont {Ghosh}}, \bibinfo {author} {\bibfnamefont {V.~K.}\ \bibnamefont {Chikkadi}}, \bibinfo {author} {\bibfnamefont {P.}~\bibnamefont {Schall}}, \bibinfo {author} {\bibfnamefont {J.}~\bibnamefont {Kurchan}}, \ and\ \bibinfo {author} {\bibfnamefont {D.}~\bibnamefont {Bonn}},\ }\href {\doibase 10.1103/PhysRevLett.104.248305} {\bibfield  {journal} {\bibinfo  {journal} {Phys. Rev. Lett.}\ }\textbf {\bibinfo {volume} {104}},\ \bibinfo {pages} {248305} (\bibinfo {year} {2010})}\BibitemShut {NoStop}%
\bibitem [{\citenamefont {Chen}\ \emph {et~al.}(2010)\citenamefont {Chen}, \citenamefont {Ellenbroek}, \citenamefont {Zhang}, \citenamefont {Chen}, \citenamefont {Yunker}, \citenamefont {Henkes}, \citenamefont {Brito}, \citenamefont {Dauchot}, \citenamefont {van Saarloos}, \citenamefont {Liu},\ and\ \citenamefont {Yodh}}]{PhysRevLett.105.025501}%
  \BibitemOpen
  \bibfield  {author} {\bibinfo {author} {\bibfnamefont {K.}~\bibnamefont {Chen}}, \bibinfo {author} {\bibfnamefont {W.~G.}\ \bibnamefont {Ellenbroek}}, \bibinfo {author} {\bibfnamefont {Z.}~\bibnamefont {Zhang}}, \bibinfo {author} {\bibfnamefont {D.~T.~N.}\ \bibnamefont {Chen}}, \bibinfo {author} {\bibfnamefont {P.~J.}\ \bibnamefont {Yunker}}, \bibinfo {author} {\bibfnamefont {S.}~\bibnamefont {Henkes}}, \bibinfo {author} {\bibfnamefont {C.}~\bibnamefont {Brito}}, \bibinfo {author} {\bibfnamefont {O.}~\bibnamefont {Dauchot}}, \bibinfo {author} {\bibfnamefont {W.}~\bibnamefont {van Saarloos}}, \bibinfo {author} {\bibfnamefont {A.~J.}\ \bibnamefont {Liu}}, \ and\ \bibinfo {author} {\bibfnamefont {A.~G.}\ \bibnamefont {Yodh}},\ }\href {\doibase 10.1103/PhysRevLett.105.025501} {\bibfield  {journal} {\bibinfo  {journal} {Phys. Rev. Lett.}\ }\textbf {\bibinfo {volume} {105}},\ \bibinfo {pages} {025501} (\bibinfo {year} {2010})}\BibitemShut {NoStop}%
\end{thebibliography}
\end{document}